\documentclass[11pt]{article}

\usepackage[
    colorlinks=true,
    linkcolor=black,    
    citecolor=black,    
    urlcolor=blue       
]{hyperref}

\usepackage{graphicx}
\usepackage{subcaption}

\usepackage{amsthm,amsmath,amssymb}
\usepackage{enumerate}
\usepackage{amssymb}

\usepackage[a4paper,margin=3 cm]{geometry}
\usepackage[numbers]{natbib}
\usepackage[normalem]{ulem}

\usepackage{enumitem}

\newlist{assumption}{enumerate}{1}
\setlist[assumption]{label=(A.\arabic*), ref=A.\arabic*}

\newtheorem{dfn}{Definition}[section]
\newtheorem{tw}[dfn]{Theorem}

\newtheorem{prop}[dfn]{Proposition}
\newtheorem{rem}[dfn]{Remark}

\newtheorem{lem}[dfn]{Lemma}

\newtheorem{cor}[dfn]{Corollary}

\numberwithin{equation}{section}

\title{Implied volatility models without static arbitrage}

\author{Micha\l \ Barski \thanks{Corresponding author: Micha\l \ Barski (mbarski@mimuw.edu.pl)}
 \\ \small  Faculty of Mathematics, Informatics and Mechanics, University of Warsaw, Poland\\
 \small{\it mbarski@mimuw.edu.pl} \bigskip \\
Maciej Wi\'sniewolski  \\ \small  Faculty of Mathematics, Informatics and Mechanics, University of Warsaw, Poland\\
 \small{\it wisniewolski@mimuw.edu.pl}}

\begin{document}

\maketitle

\begin{abstract}
We investigate the absence of static arbitrage in models describing the stochastic evolution of implied volatility surfaces. For Gaussian factor models, we derive explicit formulas for the probabilities of calendar spread, call spread and butterfly arbitrage, showing that these probabilities are strictly positive under suitable non-degeneracy conditions. We then develop a framework based on bounded stochastic factors and establish deterministic necessary and sufficient conditions for the absence of static arbitrage, together with simpler sufficient conditions. As an application, we construct an arbitrage free modification of the Cont and Vuleti\'c model using Jacobi diffusions, preserving its main empirical features while eliminating static arbitrage.
\end{abstract}


\section{Introduction}

The absence of arbitrage imposes strong restrictions on option prices
across strikes and maturities. Model independent conditions ensuring
consistency of option prices with the absence of static arbitrage were
studied, among others, by Carr and Madan \cite{CarrMadan2005SufficientNoArbitrage} and Davis and
Hobson \cite{DavisHobson2007Range}. In the case of European call options, these
conditions are expressed through monotonicity with respect to maturity
and strike and convexity with respect to strike, corresponding,
respectively, to the absence of calendar spread, call spread and
butterfly arbitrage. Their implications for the construction of implied volatility surfaces 
have subsequently been studied in a number of works, see, for example,
Roper \cite{Roper2010} and the references therein. A prominent example is the SVI parameterization introduced and developed by Gatheral; see \cite{Gatheral2006Book}. Gatheral and Jacquier \cite{gatheraljacquier2014} developed the Surface SVI (SSVI) parameterization, which provides a parsimonious construction of entire implied volatility surfaces and derived explicit conditions ensuring the absence of static arbitrage. Further extensions of this approach were developed by Guo, Jacquier, Martini and Neufcourt \cite{GuoJacquierMartiniNeufcourt2016}. Related approaches to
the estimation and construction of implied volatility surfaces can be
found, for example, in Fengler \cite{fengler2005semiparametric}.

The constructions mentioned above focus on implied volatility surfaces at a fixed observation time. They provide parameterizations satisfying the static no-arbitrage conditions across strikes and maturities, which are highly useful for fitting option prices at the observation time. They do not, however,  specify how the resulting surfaces evolve randomly through time. The relevant question is then whether the classical static no-arbitrage conditions remain satisfied almost surely at every observation time. This requirement concerns the entire evolution of the random surface, rather than a single deterministic surface, and should be distinguished from dynamic no-arbitrage conditions, which impose additional consistency requirements on option prices across different observation times. A dynamic extension of the SSVI framework was investigated by El Amrani, Jacquier and Martini \cite{ElAmraniJacquierMartini2021}, who studied stochastic dynamics of symmetric SSVI smiles for a fixed maturity, taking into account both static and dynamic no-arbitrage considerations. Their construction, however, concerns a single maturity and remains tied to a specific parametric family. A different point of view is to model directly the empirical dynamics
of the implied volatility surface. This approach was recently pursued
by Cont and Vuleti\'c \cite{ContVuletic2023Simulation}. Based on a principal component
analysis of observed implied volatility surfaces, they identify a
small number of dominant deformation patterns and model their time 
evolution by means of Gaussian factors. In particular,
Ornstein - Uhlenbeck processes provide a relevant description of the
observed features of the principal components.
The resulting model reproduces important empirical features of the
dynamics of the volatility surface. At the same time, simulated
surfaces turned out to violate the static no arbitrage restrictions, a phenomenon
which is investigated numerically in \cite{ContVuletic2023Simulation}.

The first objective of the present paper is to show that the phenomenon mentioned above
is not merely a consequence of a particular calibration or simulation
scheme. We consider a general class of implied volatility models in
which log-volatility is an affine functional of Gaussian factors. We
derive explicit representations for the probabilities of calendar
spread, call spread and butterfly arbitrage. These representations show
that, under natural non-degeneracy assumptions, the corresponding
arbitrage probabilities are strictly positive. Thus the occurrence of
static arbitrage is an intrinsic feature of the non-degenerate Gaussian
factor framework rather than a peculiarity of a particular
implementation of the Cont and Vuleti\'c model.

The second objective of this paper is to develop an alternative to Gaussian factor models that 
makes the stochastic evolution of the implied volatility surface free of static arbitrage. 
Motivated by the preceding result on the Gaussian model, we replace Gaussian factors with stochastic drivers having bounded support . The boundedness of the factors is the key structural ingredient of our approach. We derive necessary and sufficient conditions, together with simpler sufficient conditions, for the absence of static arbitrage. These conditions are deterministic and involve the deformation functions and their derivatives. Their verification admits a geometric interpretation in terms of convex polytopes. Importantly, these restrictions are separated from the precise stochastic dynamics of the factors. Once the factors are constrained to a prescribed
bounded state space, their dynamics may be selected so as to reproduce
the relevant empirical properties of the volatility surface dynamics. This gains a wide 
flexibility of the introduced model.

A further contribution of the paper is a direct modification of the Cont and Vuleti\'c model aimed at preserving its main empirical features while eliminating static arbitrage. To this end we replace the Gaussian Ornstein-Uhlenbeck factors with Jacobi diffusions (see, e.g. Larson \cite{larsson2025stochastic}, Forman and S{\o}rensen \cite{forman2008pearson}), which are bounded and  retain mean reverting behaviour and exponentially decaying autocorrelations. We then construct a four factor specification with explicit deformation functions inspired by the empirical principal components identified by Cont and Vuleti\'c. For this model, we derive explicit restrictions on the factor amplitudes ensuring the absence of static arbitrage at every time point almost surely.

In addition to this general framework, we present a simple one factor model in which a Gaussian process enters the implied volatility surface through a bounded transformation. By separating the dependence on moneyness and time to expiration, we obtain explicit sufficient conditions ensuring the absence of static arbitrage on compact domains. The construction also allows for dependence between the stochastic factor and the underlying asset price.

In this way, this paper provides an alternative to both purely static
arbitrage free parameterizations and Gaussian dynamic factor models.
Rather than specifying stochastic dynamics for the parameters of a fixed implied volatility parameterization, as in the SSVI-based approach of El Amrani, Jacquier and Martini \cite{ElAmraniJacquierMartini2021}, we model the evolution of the surface through bounded stochastic factors and impose no arbitrage restrictions on its deterministic deformation functions.
This allows the empirical dynamics and the static no arbitrage requirements to be treated as two distinct,
but compatible, components of the model.

The paper is organized as follows. Section \ref{section static arbitrage} recalls the classical static no-arbitrage conditions and reformulates them in terms of implied volatility surfaces. In Section \ref{section Gaussian implied volatility surface}, we investigate Gaussian factor models, derive explicit formulas for the probabilities of static arbitrage, and apply these results to the Cont and Vuleti\'c model. Section \ref{section Arbitrage free volatility surfaces} develops a framework for constructing arbitrage-free volatility surfaces using bounded stochastic factors. We establish deterministic necessary and sufficient conditions, together with simpler sufficient conditions, for the absence of static arbitrage, and present an additional one factor model. Finally, in Section \ref{section A static-arbitrage-free modification of the Cont-Vuletic model} we present a modification of the Cont and Vuleti\'c model based on Jacobi diffusions and construct explicit deformation functions ensuring the absence of static arbitrage. Most of the proofs are postponed to Section \ref{sec Proofs}.

\section{Static arbitrage}\label{section static arbitrage}
Denote $\mathbb{R}_+ = (0,\infty)$ and $\bar{\mathbb{R}}_+ = [0,\infty)$. 
We model the evolution of {\it implied volatility} by means of a positive random field
\[
\sigma:
\Omega\times \bar{\mathbb{R}}_+\times\mathbb{R}_+\times\mathbb{R}_+
\longrightarrow
\mathbb{R}_+,
\]
defined on a probability space $(\Omega,\mathcal F,\mathbb P)$. For every \(t\ge0\), the random field
\[
(m,v)\longmapsto \sigma_t(m,v)
\]
represents the {\it implied volatility surface} observed at time \(t\), where \(m\) denotes
moneyness and \(v\) time to expiration (also called time-to-maturity in the option
pricing literature). Throughout the paper we assume that each implied volatility surface $\sigma_t(\cdot,\cdot)$
belongs to the class
$C^{2,1}\bigl(\mathbb{R}_+^2\bigr)$
for \(\mathbb P\)-almost every \(\omega\in\Omega\).

At any fixed time \(t\), the volatility surface \(\sigma_t(m,v)\), together with the spot
value \(S_t\), determines prices of European call options through the Black--Scholes
formula
\begin{gather}\label{wzor na cene opcji}
C_t(m,v)
=
S_t \ c^{BS}\bigl(m,v,\sigma_t(m,v)\bigr),
\end{gather}
where
\[
c^{BS}(m,v,\sigma)
=
\Phi(d_1)-m\Phi(d_2)
\]
and
\begin{gather}\label{wzory na d1 i d2}
d_1
=
d_1(\sigma,m,v)
:=
\frac{-\log m+\frac12\sigma^2v}
{\sigma\sqrt v},
\qquad
d_2
=
d_2(\sigma,m,v)
:=
d_1-\sigma\sqrt v.
\end{gather}
Here \(\Phi\) denotes the cumulative distribution function of the standard normal
distribution and
\[
\varphi(x)
=
\frac{1}{\sqrt{2\pi}}e^{-x^2/2},
\qquad x\in\mathbb R,
\]
its density.

The aim of this paper is to study the evolution of volatility surfaces under the
requirement that static arbitrage is excluded at every time \(t\ge0\).

\begin{dfn}\label{def no static arbitrage}
Fix $t\geq 0$. Let $D_t$ be a Borel subset of $\mathbb{R}^2_+$. We say that there is no static arbitrage at time \(t\ge 0\) in the domain $D_t$ if and only if the following conditions hold

\begin{gather}\label{no calendar spread def}
\mathbb P
\Bigl(
\partial_v C_t(m,v)\ge0 \ 
\text{ for all }(m,v)\in D_t
\Bigr)
=
1;\\\label{no call spread def}
\mathbb P
\Bigl(
\partial_m C_t(m,v)\le0 \
\text{ for all }(m,v)\in D_t
\Bigr)
=
1;\\\label{no butterfly spread def}
\mathbb P
\Bigl(
\partial_{mm}C_t(m,v)\ge0 \ 
\text{ for all }(m,v)\in D_t
\Bigr)
=
1.
\end{gather}
In particular, if \eqref{no calendar spread def}-\eqref{no butterfly spread def} hold for
$
D_t=\{(m_0,v_0)\}
$
with some \(m_0>0\) and \(v_0>0\), then we say that there is no static arbitrage at point \((m_0,v_0)\) at time $t$.
\end{dfn}

The notion of static arbitrage adopted in this paper is based on the classical characterization of arbitrage-free option price surfaces. It is well known that, for a deterministic family of European call option prices, the absence of static arbitrage is equivalent to the absence of calendar spread, call spread and butterfly spread arbitrage. These conditions are expressed through the monotonicity of call prices with respect to maturity, the monotonicity with respect to moneyness (or equivalently strike), and the convexity with respect to moneyness. We refer to Carr  and Madan \cite{CarrMadan2005SufficientNoArbitrage}, Fengler~\cite{fengler2005semiparametric}, and Gatheral and Jacquier \cite{gatheraljacquier2014} for classical characterizations of static arbitrage in option prices and implied volatility surfaces. Since the present paper deals with random volatility surfaces, we extend the classical deterministic definition by requiring the corresponding inequalities to hold almost surely. In other words, for every fixed time \(t\), the realized option price surface is assumed to be free of static arbitrage with probability one. Condition \eqref{no calendar spread def} excludes calendar spread arbitrage, condition \eqref{no call spread def} excludes call spread arbitrage, while condition \eqref{no butterfly spread def} excludes butterfly arbitrage.

Since option prices are related to the implied volatility surface through the
Black - Scholes formula \eqref{wzor na cene opcji}, the no-arbitrage conditions from
Definition \ref{def no static arbitrage} can be reformulated directly in terms of the random field
$\sigma_t(m,v)$. The resulting conditions involve the first and second derivatives of the
volatility surface with respect to moneyness and time to expiration.
For later use, for \(\sigma>0\), \(m>0\), \(v>0\) we introduce two auxiliary functions  $H_{m,v}:\mathbb{R}\longrightarrow\mathbb R$ and a quadratic one $P_{\sigma,m,v}:\mathbb R\longrightarrow\mathbb R$
defined by

\begin{gather}\label{H funkcja definicja}
H_{m,v}(z)
:=
\frac{\Phi(d_2(e^z))}
{m e^z\sqrt v\,\varphi(d_2(e^z))},
\quad 
\text{where} \ 
d_2(e^z):=d_2(e^z,m,v), \qquad z\in\mathbb R,
\end{gather}
\begin{gather}\label{P funkcja kwadratowa}
P_{\sigma,m,v}(z)
:=
(1+d_1d_2)z^2
+
\frac{2d_1}{m\sigma\sqrt v}z
+
\frac{1}{m^2\sigma^2v}, \qquad z\in\mathbb R,
\end{gather}
where
$
d_1=d_1(\sigma,m,v),
d_2=d_2(\sigma,m,v)
$
are given by \eqref{wzory na d1 i d2}. The functions $H_{m,v}$ and $P_{\sigma,m,v}$ will be used
throughout the paper to express call and the butterfly spread
no-arbitrage conditions, respectively.

\begin{lem}\label{lem no static arbitrage}
Let \(t> 0\) and let $D_t$ be as in the Definition \ref{def no static arbitrage}. Then there is no static arbitrage at time \(t\) in the domain \(D_t\) if and only if the following conditions hold:

\medskip

\noindent
\textbf{No calendar spread arbitrage:}
\begin{gather}\label{no calendar spread}
\mathbb P\Bigg(
\partial_v \log \sigma_t(m,v)+\frac{1}{2v}
\ge0 \
\text{ for all }(m,v)\in D_t
\Bigg)
=
1.
\end{gather}

\medskip

\noindent
\textbf{No call spread arbitrage:}
\begin{gather}\label{no call spread}
\mathbb{P}\left(
\partial_m\log\sigma_t(m,v)
\leq H_{m,v}\bigl(\log\sigma_t(m,v)\bigr)
\text{ for all }(m,v)\in D_t
\right)=1.
\end{gather}
\medskip

\noindent
\textbf{No butterfly arbitrage:}
\begin{gather}\label{no butterfly spread}
\mathbb{P}\left(
\partial_{mm}\log\sigma_t(m,v)
+
P_{\sigma_t(m,v),m,v}
\bigl(\partial_m\log\sigma_t(m,v)\bigr)
\geq 0
\text{ for all }(m,v)\in D_t
\right)=1.
\end{gather}
\end{lem}
\noindent

\section{Gaussian implied volatility surface}\label{section Gaussian implied volatility surface}
The characterization established in Lemma \ref{lem no static arbitrage} is valid for an arbitrary implied
volatility field. In practice, however, it is desirable to work with more structured
models which allow for an explicit analysis of the no-arbitrage conditions.  A natural class of stochastic implied volatility models is obtained by representing the log-volatility surface in terms of Gaussian factors.
An example is the model of Cont and Vuleti\'c \cite{ContVuletic2023Simulation}, which uses a finite
collection of independent  Ornstein-Uhlenbeck processes. Motivated by this approach, we consider implied volatility surfaces of the form

\begin{gather}\label{impl vol gaussian}
\sigma_t(m,v)
=
\sigma_0(m,v)\ 
\exp\!\left(
\sum_{i\in I}X_t^{i} \ f^i(m,v)
\right), \quad t\geq0, \quad (m,v)\in\mathbb{R}_+\times\mathbb{R}_+,
\end{gather}

where:

\begin{itemize}
\item \(I\) is a finite or countable index set;

\item \(\sigma_0\) is a positive function of class
$
C^{2,1}\bigl(\mathbb{R}_+\times\mathbb{R}_+\bigr)
$
playing role of the reference volatility surface, 
\item $(f^i),  \ i\in I$,  is a sequence of deterministic real-valued functions of class
$
C^{2,1}\bigl(\mathbb{R}_+\times\mathbb{R}_+\bigr);
$

\item
\(
(X_t^{i})_{t\ge0},
\ i\in I,
\)
are independent Gaussian processes.
\end{itemize}
\noindent
Since \(\sigma_0\) is positive, it can be represented in the form
\[
\sigma_0(m,v)=e^{f^0(m,v)}
\]
for a suitable function
$
f^0\in
C^{2,1}\bigl(\mathbb{R}_+\times\mathbb{R}_+\bigr).
$
To keep the notation consistent, we assume that \(0\notin I\). Then
\[
\log \sigma_t(m,v)
=
f^0(m,v)
+
\sum_{i\in I}X_t^{i}\,f^i(m,v).
\]

Thus, for every fixed $t$, the log-volatility surface is an affine
functional of the Gaussian random variables $(X_t^i)_{i\in I}$.
A fundamental limitation of this specification is that, under natural
non-degeneracy conditions, violations of the corresponding static
no arbitrage conditions occur with positive probability. On the other
hand, the Gaussian structure makes these violations particularly
tractable, as their probabilities can be computed explicitly. The next
result gives the probabilities of calendar spread, call spread and
butterfly arbitrage at a fixed point $(m,v)$ and time $t$. For simplicity we use the notation 
$$
\partial_m f^i(m,v)=f^i_m(m,v), \quad \partial_{mm} f^i(m,v)=f^i_{mm}(m,v), \quad \partial_v f^i(m,v)=f^i_v(m,v).
$$

\begin{tw}\label{tw o prawd. statycznego arbitrazu}
Fix \(t\ge0\) and $(m,v)\in\mathbb{R}_+\times\mathbb{R}_+$. Define
\[
F_t
:=
f^0(m,v)+\sum_{i\in I}X_t^{i}\ f^i(m,v),\qquad
M_t
:= f^0_m(m,v)+\sum_{i\in I}X_t^{i}\ f^i_m(m,v),
\]
\[
K_t
:=
f^0_{mm}(m,v)+\sum_{i\in I}X_t^{i}\ f^i_{mm}(m,v),\qquad 
V_t
:=
 f^0_v(m,v)+\sum_{i\in I}X_t^{i}\ f^i_v(m,v).
\]

Assume that the above random variables are well-defined and Gaussian. Recall, the functions 
$H_{m,v}$ and $P_{\sigma,m,v}$ are given by 
\eqref{H funkcja definicja} and \eqref{P funkcja kwadratowa}.

\begin{enumerate}
\item[(i)]
If
$
\operatorname{Var}(V_t)>0,
$
then the probability of calendar spread arbitrage at the point \((m,v)\) and
time \(t\) is equal to
\begin{gather}\label{prawd. calendar gaussian}
\Phi\left(
\frac{
-\frac{1}{2v}-\mathbb E[V_t]
}
{
\sqrt{\operatorname{Var}(V_t)}
}
\right).
\end{gather}

\item[(ii)]
Assume that
$
\operatorname{Var}(F_t)>0
$
and that the conditional variance of \(M_t\) given \(F_t\) is positive.
Then the probability of call spread arbitrage at the point \((m,v)\) and time
\(t\) is equal to
\begin{gather}\label{prawd. call gaussian}
\int_{\mathbb R}
\left[
1-
\Phi\left(
\frac{
H_{m,v}(y)-\mu_{M|F}(y)
}
{
\sigma_{M|F}
}
\right)
\right]
f_F(y)\,dy,
\end{gather}
where \(f_F\) is the density of \(F_t\), and
\[
\mu_{M|F}(y)
:=
\mathbb E[M_t]
+
\frac{\operatorname{Cov}(M_t,F_t)}
{\operatorname{Var}(F_t)}
\bigl(y-\mathbb E[F_t]\bigr),
\qquad
\sigma_{M|F}^2
:=
\operatorname{Var}(M_t)
-
\frac{\operatorname{Cov}(M_t,F_t)^2}
{\operatorname{Var}(F_t)}.
\]

\item[(iii)]
Assume that the conditional variance of \(K_t\) given \((F_t,M_t)\)
is positive. Then the probability of butterfly arbitrage at the point
\((m,v)\) and time \(t\) is equal to
\begin{gather}\label{prawd. butterfly gaussian}
\int_{\mathbb R}\int_{\mathbb R}
\Phi\left(
\frac{
-P_{e^y,m,v}(z)-\mu_{K|F,M}(y,z)
}
{
\sigma_{K|F,M}
}
\right)
f_{M|F=y}(z)f_F(y)\,dz\,dy,
\end{gather}
where \(f_F\) is the density of \(F_t\), \(f_{M|F=y}\) is the conditional density of
\(M_t\) given \(F_t=y\), and
\[
\mathcal{L}(K_t\,\big|\,F_t=y,M_t=z)
=
N\left(
\mu_{K|F,M}(y,z),
\sigma_{K|F,M}^2
\right).
\]
\end{enumerate}
\end{tw}

The formulas \eqref{prawd. calendar gaussian},\eqref{prawd. call gaussian} and \eqref{prawd. butterfly gaussian}  in Theorem \ref{tw o prawd. statycznego arbitrazu} show, in particular, that under the
corresponding non-degeneracy assumptions the probabilities of calendar
spread, call spread and butterfly arbitrage at a fixed point are strictly
positive. Thus, in the non-degenerate Gaussian setting, the 
static no arbitrage conditions cannot hold almost surely.

As the straightforward consequence of the proof of Theorem \ref{tw o prawd. statycznego arbitrazu} we obtain
\begin{cor} Let $t>0$ and $\Delta\subset\mathbb{R}^d$ be compact. The same arguments used to establish Theorem \ref{tw o prawd. statycznego arbitrazu} work if we replace $F_t$ by
\begin{align}\label{white noise representation}
	F_t = f^0(m,v)+ \int_{[0,t]\times \Delta}h_{m,v}(s,x)B(ds,dx),
\end{align}
where 
\begin{enumerate}
\item{} $B$ is a white noise on $[0,t]\times \Delta$; 
\item{} for each fixed $m,v$ we have $h_{m,v} \in L^2([0,t]\times \Delta)$;
\item{} for each fixed $(s,x) \in [0,t]\times \Delta$ the function $(m,v) \mapsto h_{m,v}(s,x)$ is measurable.
\end{enumerate}
Indeed, for every fixed $(m,v)$ the stochastic integral in \eqref{white noise representation} is a Gaussian random variable. Consequently, the idea of the proof of Theorem \ref{tw o prawd. statycznego arbitrazu} remains unchanged and its main argument relies on the Gaussianity of the random vector involved.
\end{cor}

\subsection{The model of Cont and Vuleti\'c}

The model of Cont and Vuleti\'c is obtained as a particular case of the Gaussian
framework based on a finite number of independent Ornstein - Uhlenbeck factors. More precisely, we consider now the model of the form
\[
\sigma_t(m,v)
=
\exp\left(
f^0(m,v)+\sum_{i=1}^n X_t^{i}\ f^i(m,v)
\right),
\]
where
\[
dX_t^{i}
=
\lambda_i(\alpha_i-X_t^{i})\,dt+\gamma_i\,dW_t^i, \quad X_0^{i} = x_0^{i}, 
\qquad i=1,\ldots,n,
\]
where $W = (W^{1},..,W^{n})$ is $n$-dimensional Wiener process.
The explicit form of the Ornstein--Uhlenbeck factors gives
\[
X_t^{i}
=
\alpha_i+(X_0^{i}-\alpha_i)e^{-\lambda_i t}
+
\gamma_i\int_0^t e^{-\lambda_i(t-s)}\,dW_s^i, \qquad i=1,2,...,n.
\]
Hence
\[
X_t^{i}\sim N(\mu_i(t),s_i^2(t)),
\]
where
\[
\mu_i(t)
=
\alpha_i+(x_0^{i}-\alpha_i)e^{-\lambda_i t},
\qquad
s_i^2(t)
=
\frac{\gamma_i^2}{2\lambda_i}
\left(1-e^{-2\lambda_i t}\right).
\]

Since an Ornstein - Uhlenbeck process is a Gaussian process and its explicit
mean and covariance structure are known, the probabilities of violation arbitrage conditions from Theorem \ref{tw o prawd. statycznego arbitrazu} can be written in a straightforward way.

\begin{cor}[Static arbitrage probabilities in the Cont--Vuleti\'c model]

Fix \(t\ge0\) and \((m,v)\in\mathbb{R}_+\times\mathbb{R}_+\). For a deterministic
function \(g\), write \(g=g(m,v)\). Define
\[
\mu_F
=
f^0+\sum_{i=1}^n \mu_i(t)\ f^i,
\qquad
\mu_M
= f^0_m+\sum_{i=1}^n \mu_i(t) f^i_m,
\]
\[
\mu_K
=
f^0_{mm}+\sum_{i=1}^n \mu_i(t)f^i_{mm},
\qquad
\mu_V
=
f^0_v+\sum_{i=1}^n \mu_i(t) f^i_v.
\]
For \(A,B\in\{F,M,K,V\}\), define
\[
\Sigma_{AB}(t)
:=
\sum_{i=1}^n g_A^i g_B^i\,s_i^2(t),
\]
where
$
g_F^i=f^i, g_M^i= f^i_m, g_K^i=f^i_{mm}, g_V^i=f^i_v.
$
Then the following statements are true.

\begin{enumerate}
\item[(i)] The probability of calendar spread arbitrage at the point \((m,v)\) and
time \(t\) is equal to
\[
\Phi\left(
\frac{-\frac1{2v}-\mu_V}
{\sqrt{\Sigma_{VV}(t)}}
\right),
\]
provided \(\Sigma_{VV}(t)>0\).
\item[(ii)]
If
\[
\Sigma_{FF}(t)>0
\quad\text{and}\quad
\Sigma_{MM}(t)-\frac{\Sigma_{MF}(t)^2}{\Sigma_{FF}(t)}>0,
\]
then the probability of call spread arbitrage at the point \((m,v)\) and time
\(t\) is equal to
\[
\int_{\mathbb R}
\left[
1-
\Phi\left(
\frac{H_{m,v}(y)-\mu_{M|F}(y)}
{\sigma_{M|F}}
\right)
\right]
f_F(y)\,dy,
\]
where
\[
\mu_{M|F}(y)
=
\mu_M
+
\frac{\Sigma_{MF}(t)}{\Sigma_{FF}(t)}(y-\mu_F),
\]
and
\[
\sigma_{M|F}^2
=
\Sigma_{MM}(t)-\frac{\Sigma_{MF}(t)^2}{\Sigma_{FF}(t)}.
\]
\item[(iii)]
Assume that the conditional variance of \(K_t\) given \((F_t,M_t)\) is positive.
For 
\[
U:=
\begin{pmatrix}
F\\ M
\end{pmatrix}.
\]
denote
\[
\Sigma_{UU}(t)
=
\begin{pmatrix}
\Sigma_{FF}(t)&\Sigma_{FM}(t)\\
\Sigma_{MF}(t)&\Sigma_{MM}(t)
\end{pmatrix},
\qquad
\Sigma_{KU}(t)
=
\begin{pmatrix}
\Sigma_{KF}(t)&\Sigma_{KM}(t)
\end{pmatrix},
\]
and assume that $\Sigma_{UU}(t)$ is positive definite.
Then
\[
\mu_{K|F,M}(y,z)
=
\mu_K
+
\Sigma_{KU}(t)\Sigma_{UU}(t)^{-1}
\left(
\begin{pmatrix}
y\\ z
\end{pmatrix}
-
\begin{pmatrix}
\mu_F\\ \mu_M
\end{pmatrix}
\right),
\]
and
\[
\sigma_{K|F,M}^2
=
\Sigma_{KK}(t)
-
\Sigma_{KU}(t)\Sigma_{UU}(t)^{-1}\Sigma_{UK}(t).
\]
Then the probability of butterfly arbitrage at the point
\((m,v)\) and time \(t\) is equal to
\[
\int_{\mathbb R}\int_{\mathbb R}
\Phi\left(
\frac{
-P_{e^y,m,v}(z)-\mu_{K|F,M}(y,z)}
{\sigma_{K|F,M}}
\right)
f_{M|F=y}(z)f_F(y)\,dz\,dy.
\]
\end{enumerate}
\end{cor}

\section{Arbitrage free volatility surfaces}\label{section Arbitrage free volatility surfaces}
The results of the previous section show that Gaussian volatility models provide a tractable framework for the analysis of static arbitrage. In particular, the probabilities of calendar spread, call spread and butterfly arbitrage can be expressed explicitly in terms of finite-dimensional Gaussian distributions. At the same time, the Gaussian framework possesses an intrinsic limitation. Since Gaussian random variables have unbounded support, the corresponding volatility surfaces may attain arbitrarily large or arbitrarily small values with positive probability. Consequently, the no-arbitrage conditions derived in Lemma \ref{lem no static arbitrage} cannot be enforced by restricting the parameters of the model. Instead, one can quantify the probability of arbitrage.

This observation motivates the search for alternative stochastic drivers leading to
arbitrage-free volatility surfaces. In this section we present two different approaches.
The first one is based on bounded stochastic factors and yields a geometric characterization
of the absence of static arbitrage together with explicit necessary and sufficient conditions.
As an application, it provides a direct arbitrage-free refinement of the Cont-Vuleti\'c model
obtained by replacing the Ornstein-Uhlenbeck factors with Jacobi diffusions. The second
approach is a tractable one-factor construction in which the Gaussian driver is transformed
through a bounded function and the dependence on moneyness and time to expiration is separated. 
This makes it possible to derive explicit conditions excluding all three types of static arbitrage while preserving a simple
probabilistic structure.

\subsection{Implied volatility beyond Gaussian framework}
 In this section we consider implied volatility surfaces of the form 
\begin{gather}\label{model with bounded factors 1} 
\sigma_t(m,v) = \exp\!\left( f^0(m,v)+\sum_{i=1}^{n} Y_t^i\cdot f^i(m,v) \right), 
\end{gather}
where \(f^0,f^1,\ldots,f^n\) are deterministic functions and \((Y^i)_{i=1,\ldots,n}\) are independent processes satisfying 
\begin{gather}\label{model with bounded factors 2}
\mathbb P\!\left(
Y_t^i\in[0,1]
\text{ for all } t\ge0
\right)=1.
\end{gather}

We additionally assume that, for every  $t>0$ and  $i=1,…,n$ the topological support of the distribution of $Y_t^i$
is the entire interval  $[0,1]$. The full support assumption can be relaxed, but it simplifies the proofs of the results below.
By independence of the factors we get for $Y_t:=(Y^1_t,...,Y^n_t)$ that 
\begin{gather}\label{full support}
{\rm supp} \ \mathcal{L}(Y_t)=[0,1]^n,\qquad t>0.
\end{gather}

 An example which is of particular interest in our setting is the Jacobi process discussed in Section \ref{Section noise adjustment}. For the Jacobi process living in $[0,1]$ it is known from the general theory that the boundary points are unattainable, while the transition density is strictly positive on $(0,1)$. Consequently, its distribution has full support $[0,1]$ for $t>0$.

  Restricting the support of the driving process is, probably, not the only possible mechanism for eliminating static arbitrage. Nevertheless, it leads to a particularly transparent geometric characterization of the no-arbitrage conditions. Here we derive explicit necessary and sufficient conditions for the absence of static arbitrage and also formulate simpler sufficient conditions, which are easy to implement. The analysis will be carried out separately for calendar spread, call spread and butterfly arbitrage.

\subsubsection{Calendar spread arbitrage for bounded factors}

The calendar spread condition is the simplest among the three types of static arbitrage. The affine dependence on the bounded stochastic factors allows us to derive an explicit necessary and sufficient condition for its absence.

\begin{prop}\label{proposition Jacobi calendar}
The model \eqref{model with bounded factors 1}-\eqref{model with bounded factors 2} is free of calendar spread arbitrage at the
point \((m,v)\) and any time \(t\geq 0\) if and only if
\[
\sum_{i=1}^n \bigl(f_v^i(m,v)\bigr)^-
\le
\frac1{2v}+f_v^0(m,v).
\]
\end{prop}

\subsubsection{Call spread arbitrage for bounded factors}

The analysis of call spread arbitrage is more involved. Although the
corresponding condition is no longer affine, it still depends on the
factors through two affine functionals. This makes it possible to reformulate
the problem geometrically in terms of a convex polygon in $\mathbb{R}^2$.  As a result, the
verification of the no call spread arbitrage condition reduces to a finite collection of
one dimensional tests performed along the upper boundary of this polygon. Hence the problem becomes computationally tractable in practical applications. 

\begin{prop}
\label{prop no call spread Jacobi}
Fix \((m,v)\) and define
\[
a_i:=f^i(m,v),
\qquad
b_i:=f^i_m(m,v),
\qquad i=1,\ldots,n,
\]
$a:=(a_1,...,a_n), b:=(b_1,...,b_n)$ together with
\[
c:=f^0(m,v),
\qquad
d:=f^0_m(m,v).
\]
Assume that $a\neq 0$ and for every vertex $u\in\{0,1\}^n$ of the cube $[0,1]^n$ define
\[
A_u:=\langle a,u\rangle,
\qquad
B_u:=\langle b,u\rangle,
\]
and let
\[
\mathcal{P}:=\operatorname{conv}
\{(A_u,B_u):u\in\{0,1\}^n\},
\]
be the corresponding polygon in $\mathbb{R}^2$. Define
\[
B(z):=
\max\{w:(z,w)\in \mathcal{P}\}.
\]
Let $(z_1,w_1),\ldots,(z_N,w_N)$, with $N\leq 2^n$, be the vertices of the upper boundary $\Gamma$ of $\mathcal{P}$:
\[
\Gamma:=\{(z,w)\in \mathcal{P}:w=B(z)\},
\]
ordered so that $z_1<\cdots<z_N$. For \(j=1,\ldots,N-1\), define
\[
\beta_j:=
\frac{w_{j+1}-w_j}{z_{j+1}-z_j},
\qquad
\alpha_j:=w_j-\beta_j z_j.
\]

Then the model \eqref{model with bounded factors 1}-\eqref{model with bounded factors 2} is free of call spread arbitrage at the point
\((m,v)\) for every \(t>0\) if and only if
\[
H_{m,v}(c+z_j)-d-w_j\ge0,
\qquad j=1,\ldots,N,
\]
and, for every \(j=1,\ldots,N-1\), every solution of the equation
\[
H^\prime_{m,v}(c+z)=\beta_j,
\qquad
z\in(z_j,z_{j+1}),
\]
satisfies
\[
H_{m,v}(c+z)-d-\alpha_j-\beta_j z\ge0.
\]

If \(a=0\), then the corresponding necessary and sufficient condition for no call spread arbitrage is
\[
H_{m,v}(c)-d-\sum_{i=1}^n b_i^+\ge0.
\]
\end{prop}

\noindent
\vskip2ex
In practical applications it is convenient to replace the previous criterion
by a simpler sufficient condition.

\begin{prop}
\label{prop suff cond for no call arb Jacobi}
Assume that
\[
H_{m,v}\left(
f^0(m,v)
+
\sum_{i=1}^n
\bigl(f^i(m,v)\bigr)^+
\right)
\ge
f^0_m(m,v)
+
\sum_{i=1}^n
\bigl(f^i_m(m,v)\bigr)^+ .
\]

Then the model \eqref{model with bounded factors 1}-\eqref{model with bounded factors 2} is free of call spread arbitrage at the point \((m,v)\) for any
\(t\geq 0\).
\end{prop}

\subsubsection{Butterfly arbitrage for bounded factors}

Among the three types of static arbitrage, the butterfly condition is the most
challenging. It involves second-order derivatives of the implied volatility and
cannot be reduced to a one-dimensional optimization problem. Nevertheless, it is possible to
associate the problem with a convex polytope in $\mathbb{R}^3$. The absence of butterfly arbitrage can then be characterized by a finite collection of conditions on a suitable triangulation of the projection of the polytope onto the first two coordinates.

\begin{prop}
\label{prop no butterfy arb Jacobi}
Fix \((m,v)\) and define
\[
a_i:=f^i(m,v), \qquad
b_i:=f^i_m(m,v), \qquad
r_i:=f^i_{mm}(m,v),
\qquad i=1,\ldots,n,
\]
$a:=(a_1,...,a_n), b:=(b_1,...,b_n), r:=(r_1,...,r_n)$ together with
\[
c:=f^0(m,v), \qquad
d:=f^0_m(m,v), \qquad
e:=f^0_{mm}(m,v).
\]
For every vertex \(q\in\{0,1\}^n\) of the cube $[0,1]^n$ set
\[
A_q:=\langle a,q\rangle,
\qquad
B_q:=\langle b,q\rangle,
\qquad
C_q:=\langle r,q\rangle,
\]
and define the convex polytope
\begin{gather}\label{def Q polytope}
Q:=
\operatorname{conv}
\left\{
(A_q,B_q,C_q):
q\in\{0,1\}^n
\right\}
\subset\mathbb R^3.
\end{gather}

Let \(D\) be the projection of \(Q\) onto the first two coordinates, that is
\[
D:=
\left\{
(z,w)\in\mathbb R^2:
\text{there exists } u\in\mathbb R
\text{ such that } (z,w,u)\in Q
\right\}.
\]

The lower boundary of $Q$ is described by the function $L:D\to\mathbb \mathbb{R}$ defined by
\[
L(z,w):=\min\{u\in\mathbb R:(z,w,u)\in Q\}.
\]

Assume that the vectors $a$ and $b$ are linearly independent.
Then $D$ has nonempty interior, and there exists a triangulation 
\[
D=\bigcup_{k=1}^N T_k
\]
with some finite $N$,  such that \(L\) is affine on each triangle \(T_k\). Thus, for
each \(k=1,\ldots,N\), there exist coefficients
$
\alpha_k,\beta_k,\gamma_k\in\mathbb R
$
such that
\[
L(z,w)=\alpha_k+\beta_k z+\gamma_k w,
\qquad
(z,w)\in T_k.
\]

Then the model \eqref{model with bounded factors 1}-\eqref{model with bounded factors 2}
is free of butterfly arbitrage at the point
\((m,v)\) and any time \(t\ge0\) if and only if, for every \(k=1,\ldots,N\),
\[
e+\alpha_k+\beta_k z+\gamma_k w
+
P_{e^{c+z},m,v}(d+w)
\ge0
\]
for all
$
(z,w)\in T_k.
$
\end{prop}

\vskip2ex

\begin{rem}
The coefficients \(\alpha_k,\beta_k,\gamma_k\) appearing in Proposition \ref{prop no butterfy arb Jacobi}
are determined explicitly by the vertices of the triangle \(T_k\). Indeed,
assume that
\[
T_k=\operatorname{conv}\{(z_1,w_1),(z_2,w_2),(z_3,w_3)\}
\]
and that the corresponding points on the lower boundary of \(Q\) are
\[
(z_1,w_1,u_1),\qquad
(z_2,w_2,u_2),\qquad
(z_3,w_3,u_3).
\]
Since \(L\) is affine on \(T_k\), we have
\[
L(z,w)=\alpha_k+\beta_k z+\gamma_k w,
\qquad (z,w)\in T_k.
\]
Therefore \(\alpha_k,\beta_k,\gamma_k\) are the unique solution of the linear system
\[
\begin{pmatrix}
1 & z_1 & w_1\\
1 & z_2 & w_2\\
1 & z_3 & w_3
\end{pmatrix}
\begin{pmatrix}
\alpha_k\\
\beta_k\\
\gamma_k
\end{pmatrix}
=
\begin{pmatrix}
u_1\\
u_2\\
u_3
\end{pmatrix}.
\]
Equivalently, they are the coefficients of the plane containing the face of
the lower boundary of \(Q\) lying above \(T_k\). Thus, once the lower boundary
of \(Q\) has been triangulated, all coefficients entering Proposition \ref{prop no butterfy arb Jacobi} are
obtained by solving a finite number of \(3\times3\) linear systems.
\end{rem}

\vskip2ex
Proposition \ref{prop no butterfy arb Jacobi} provides an exact characterization of the no-butterfly arbitrage, which 
requires checking the inequality on the entire set \(D\). The following
result gives a simple sufficient condition.

\begin{prop}
\label{Jacobi sufficient butterfly}
Assume that
\[
e-\sum_{i=1}^n r_i^-
+
\inf_{\substack{
u\in
[c-\sum_i a_i^-,
 \;c+\sum_i a_i^+]\\
w\in
[d-\sum_i b_i^-,
 \;d+\sum_i b_i^+]
}}
P_{e^u,m,\tau}(w)
\ge0,
\]
where we use the notation from Proposition \ref{prop no butterfy arb Jacobi}.
Then the model \eqref{model with bounded factors 1}-\eqref{model with bounded factors 2} is free of butterfly arbitrage at point \((m,v)\) for any $t\geq 0$.
\end{prop}

\subsection{A one factor model}
We introduce a tractable class of random volatility surfaces for which
the static no-arbitrage conditions can be verified explicitly. The main idea is to perturb a deterministic
volatility surface by a bounded nonnegative transformation of the Gaussian factor.
Boundedness is useful because it allows the butterfly
condition to be controlled uniformly on compact domains.

Let $(X_t)_{t\ge0}$ be a Gaussian process and define
\[
Y_t:=F(X_t),
\]
where $F$ is a measurable, $[0,1]$-valued function. Then
\[
0<Y_t<1
\]
almost surely. We consider the model
\begin{equation}\label{eq:one-factor-model}
\sigma_t(m,v)
=
\exp\left\{
a(v)-\lambda\rho(m)+r(v)Y_t
\right\},\qquad t\geq 0,
\qquad (m,v)\in D_t \subseteq \mathbb{R}_{+}\times\mathbb{R}_{+},
\end{equation}
where $\lambda\ge0$ and the functions
\[
a,r:(0,\infty)\to\mathbb R,
\qquad
\rho:(0,\infty)\to\mathbb R
\]
are twice continuously differentiable. The term $a(v)$ controls the deterministic term-structure component, the
function $r(v)$ controls the strength of the random factor, and the function
$\rho(m)$ introduces dependence on moneyness. The parameter $\lambda$
controls the size of the moneyness effect. The set $D_t$ is the domain on which the static arbitrage will be excluded.

\begin{prop}\label{PP}
Fix $t\geq 0$ and  let $D_t\subset(0,\infty)\times(0,\infty)$ be compact. Suppose that
\[
a'(v)\ge0,\qquad r(v)\ge0,\qquad r'(v)\ge0,
\]
and
\[
\rho'(m)\ge0,\qquad \rho''(m)\le0
\]
for all $(m,v)\in A_t$. Then there exists $\lambda_0>0$ such that for every
$\lambda\in[0,\lambda_0]$, the volatility surface defined by
\eqref{eq:one-factor-model} satisfies the no calendar spread, no call spread,
and no butterfly arbitrage conditions on $D_t$ almost surely.
\end{prop}

The construction presented above is purely pathwise and does not require any
assumption on the joint distribution of the spot price and the random factor.
A natural probabilistic specification is obtained by modelling both quantities
through a jointly Gaussian system.

Let $(V_t,X_t)_{t\ge0}$ be a two-dimensional Gaussian process defined on a
filtered probability space
$(\Omega,\mathcal F,(\mathcal F_t)_{t\ge0},\mathbb P)$.
We allow the components $V$ and $X$ to be correlated. Let
$\mu:[0,\infty)\to\mathbb R$ be a deterministic function and define
\[
S_t
=
\exp\{V_t+\mu(t)\},
\qquad t\ge0.
\]
Consequently,
\[
m
=
\frac{K}{S_t}
=
K\exp\{-V_t-\mu(t)\}.
\]

The implied volatility surface is then given by
\[
\sigma_t(m,v)
=
\exp\left\{
a(v)-\lambda\rho(m)+r(v)F(X_t)
\right\},
\]
or, equivalently, as a function of the strike,
\[
\sigma_t(K,v)
=
\exp\left\{
a(v)
-
\lambda
\rho\!\left(
K e^{-V_t-\mu(t)}
\right)
+
r(v)F(X_t)
\right\}.
\]

Finally, the call option price is defined through the Black--Scholes formula,
namely
\[
C_t(K,v)
=
S_t\,
c^{BS}
\left(
\frac{K}{S_t},
v,
\sigma_t\!\left(\frac{K}{S_t},v\right)
\right).
\]

Notice that the spot price enters the implied volatility surface only through
the moneyness variable $m=K/S_t$. The dependence between the spot price and
the volatility surface is therefore entirely determined by the joint
distribution of the Gaussian processes $(V_t,X_t)$. In particular, the
correlation structure between $V$ and $X$ may be chosen freely to model
different leverage effects.

Moreover, for every fixed time $t$, the values $S_t$ and $X_t$ are frozen,
and the static arbitrage conditions are verified pathwise with respect to the
variables $(m,v)$. Consequently, the validity of the no calendar spread, no
call spread and no butterfly arbitrage conditions established in Proposition
\ref{PP} is unaffected by the dependence structure between $V$ and $X$.
Only the probabilistic behaviour of the volatility surface changes.

\section{A static-arbitrage-free modification of the Cont-Vuleti\'c model}\label{section A static-arbitrage-free modification of the Cont-Vuletic model}

We propose a modification of the Cont and Vuleti\'c model along two complementary directions. First, we replace the Gaussian Ornstein-Uhlenbeck factors with bounded Jacobi diffusions, preserving their mean-reverting behaviour and exponential autocorrelation structure. Second, we construct deterministic deformation functions inspired by the empirical principal components and derive explicit restrictions on their amplitudes ensuring the absence of static arbitrage.

\subsection{Bounded factor dynamics}\label{Section noise adjustment}

The empirical study of Cont and Vuleti\'c indicates that the principal component factors $(X_t^i)$, $i=1,\ldots,n$,
driving the implied volatility surface \eqref{impl vol gaussian} exhibit several characteristic features. In particular, the
factors display strong positive autocorrelation, mean-reverting behaviour, and autocorrelation functions decaying approximately exponentially. Moreover, when sampled on a regular time grid, they exhibit a dependence
structure similar to that of an autoregressive process of order one. These observations motivated Cont and
Vuleti\'c to model the volatility factors by Ornstein-Uhlenbeck processes. In the previous section we showed, 
however, that Gaussian volatility models necessarily generate static arbitrage with positive probability. We therefore propose alternative factor dynamics preserving the above statistical properties while possessing a bounded state space.
The Jacobi diffusion provides such a framework.

Consider the Jacobi diffusion
\begin{equation}
dY_t=\kappa(\theta-Y_t)\,dt
    +\eta\sqrt{Y_t(1-Y_t)}\,dW_t,
\qquad Y_0\in(0,1),
\end{equation}
where
\[
\kappa>0,\qquad \theta\in(0,1),\qquad \eta>0,
\]
and
\[
2\kappa\theta>\eta^2,
\qquad
2\kappa(1-\theta)>\eta^2.
\]
The Jacobi diffusion belongs to the class of Pearson diffusions. Its properties are well documented in the
literature; see, in particular, Forman and S{\o}rensen \cite{forman2008pearson} and 
Larson \cite{larsson2025stochastic}. Under the above assumptions, the process has a unique
strong solution, and the boundary points $0$ and $1$
are unattainable, i.e.
\[
\mathbb{P}\bigl(Y_t\in(0,1)
\text{ for all }t\geq 0\bigr)=1.
\]
Moreover, the process is ergodic and admits the
invariant distribution
\[
\operatorname{Beta}(\alpha,\beta),
\qquad \text{with} \ \qquad 
\alpha=\frac{2\kappa\theta}{\eta^2},
\qquad
\beta=\frac{2\kappa(1-\theta)}{\eta^2}.
\]
As a bounded process, $Y$ has moments of any order. 
Its mean-reverting property is expressed by
\begin{equation}
\mathbb{E}[Y_{t+h}\mid\mathcal{F}_t]
=\theta+(Y_t-\theta)e^{-\kappa h},
\qquad h\geq 0.
\end{equation}
and its covariance function satisfies
\begin{equation}
\operatorname{Cov}(Y_t,Y_{t+h})
=e^{-\kappa h}\operatorname{Var}(Y_t),
\qquad t,h\geq 0.
\end{equation}
Consequently, for $t>0$,
\begin{equation}
\operatorname{Corr}(Y_t,Y_{t+h})
=e^{-\kappa h}
\sqrt{\frac{\operatorname{Var}(Y_t)}
{\operatorname{Var}(Y_{t+h})}}.
\end{equation}
Since the distribution of $Y_t$ converges to the $\operatorname{Beta}(\alpha,\beta)$, we have
\[
\operatorname{Var}(Y_t)\underset{t\to\infty}{\longrightarrow}
v_\infty
:=\frac{\alpha\beta}
{(\alpha+\beta)^2(\alpha+\beta+1)}
>0.
\]
It follows that, for every fixed $t>0$,
\begin{equation}
\operatorname{Corr}(Y_t,Y_{t+h})
\sim
\sqrt{\frac{\operatorname{Var}(Y_t)}{v_\infty}}
\,e^{-\kappa h},
\qquad h\to\infty.
\end{equation}
Thus, the autocorrelation decays asymptotically at the exponential rate $\kappa$.  When sampled at equally
spaced times, the process exhibits a corresponding geometric decay of correlations, as in an autoregressive process of order $1$, i.e. AR(1). Note that the rate of convergence $\kappa$ is the same as for the Ornstein-Uhlenbeck process.
As in the Ornstein-Uhlenbeck case, the Jacobi diffusion becomes stationary when initialized with its invariant distribution. In this regime, both processes have the same exponential autocorrelation function, although their invariant distributions are different - Gaussian for the Ornstein-Uhlenbeck process and Beta for the Jacobi diffusion.

Concluding, the Jacobi diffusion shares the mean-reverting behaviour and exponential autocorrelation structure
of the Ornstein-Uhlenbeck process, while having a bounded state space. This makes it suitable for 
constructing implied volatility models satisfying the static no arbitrage conditions developed in 
Section \ref{section Arbitrage free volatility surfaces}.

\subsection{Arbitrage-free deformation functions}
We now consider a four-factor specification of the implied
volatility surface motivated by the empirical principal component
analysis of Cont and Vuleti\'c~\cite{ContVuletic2023Simulation}.
Their results suggest that a substantial part of the variability
of implied volatility surfaces can be described by four
characteristic deformation patterns, commonly interpreted as
changes in the volatility level, skew, term structure, and
curvature.

Our objective is not to reproduce the empirical principal
components exactly. Instead, we seek simple deterministic
functions that capture their qualitative features and can be
incorporated into a model satisfying the static no-arbitrage
conditions derived in Section \ref{section Arbitrage free volatility surfaces}.

We work on a time-independent domain $D_t\equiv D$, $t\geq 0$, where
\[
D=\{(m,v):\, 0.6\leq m\leq 1.4,\quad 30/365\leq v\leq 1\}.
\]
Inspection of the four principal components defined on $D$ displayed in
Figure~4 of~\cite{ContVuletic2023Simulation} motivates the following
requirements:
\begin{itemize}
\item The first component, associated with changes in the overall
volatility level, should be positive throughout the domain.
Its magnitude should decrease with both moneyness and maturity, reflecting the downward slopes visible in the empirical
component. We capture these features by combining an exponentially decreasing term-structure profile with a positive
affine function of moneyness.

\item The second component should describe changes in the volatility
skew. It should therefore take opposite signs on the two sides
of at-the-money moneyness and vanish at $m=1$.
We also allow its magnitude to decrease with maturity.

\item The third component should primarily describe changes in the
term structure. In particular, it should change sign as
maturity increases, allowing short- and long-dated implied
volatilities to move in opposite directions. A mild dependence
on moneyness is included to avoid a completely flat profile
in that direction.

\item The fourth component should capture changes in the
curvature of the volatility smile. We seek a function that is
negative near at-the-money moneyness and positive in the wings.
A small linear term allows for asymmetry between the two wings.
\end{itemize}

\begin{figure}[htbp]
\centering

\begin{minipage}[c]{0.43 \textwidth}
    \centering
    \includegraphics[
    width=\linewidth,
]{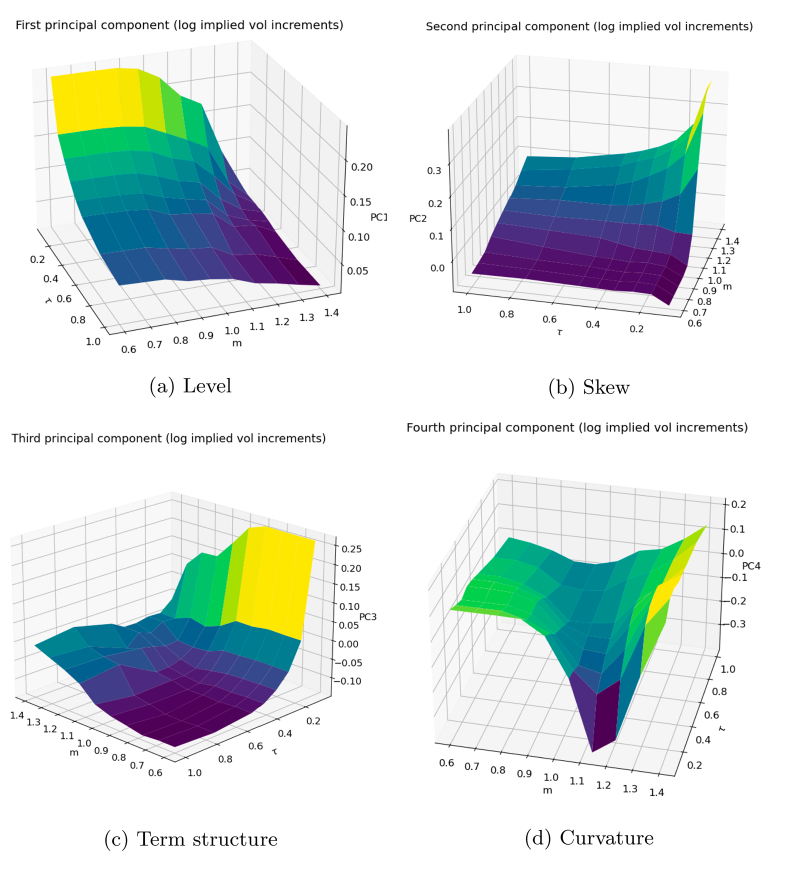}

    \smallskip
    \small (a) Empirical principal components reproduced from Cont and Vuleti\'c \cite{ContVuletic2023Simulation}. 
\end{minipage}
\hfill
\begin{minipage}[c]{0.43\textwidth}
    \centering
    \includegraphics[width=\linewidth]{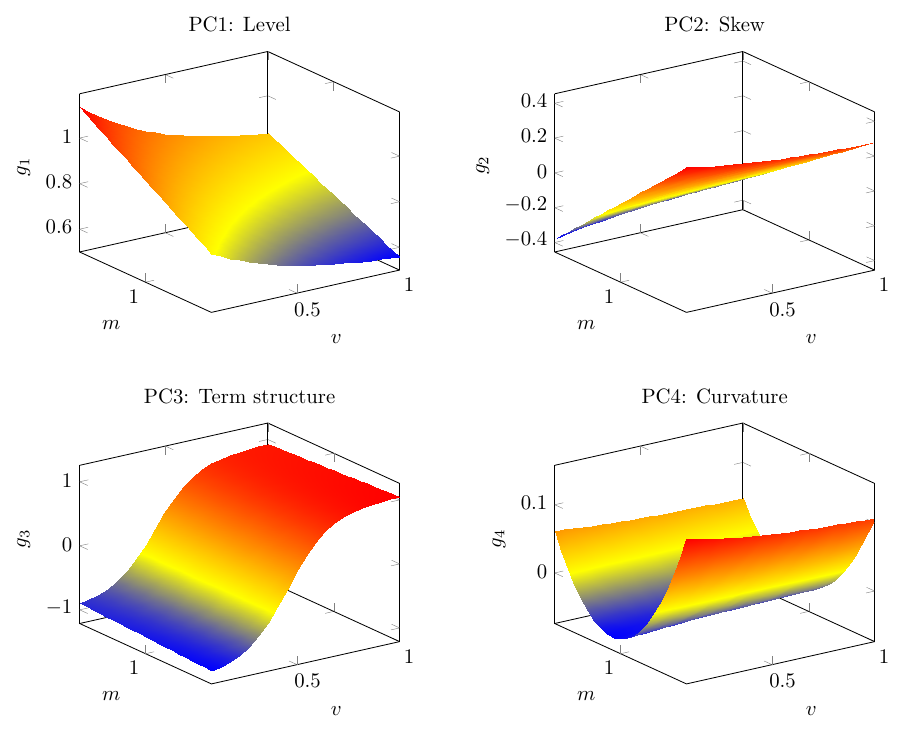}

    \smallskip
    \small (b) Our deterministic factor functions.
\end{minipage}

\caption{Comparison of the empirical principal
components of Cont and Vuleti\'c with our
deterministic factor functions.}
\label{fig:factor-comparison}

\end{figure}

These considerations lead us to propose the following
functions:
\begin{equation}\label{eq:four-factors}
\begin{aligned}
f^1(m,v)
&=(0.65+0.35e^{-2v})
  \bigl(1-0.5(m-1)\bigr),\\
f^2(m,v)
&=(m-1)(0.65+0.35e^{-2v}),\\
f^3(m,v)
&=\tanh\bigl(5(v-0.45)\bigr)
  \bigl(1+0.15(m-1)\bigr),\\
f^4(m,v)
&=\bigl((m-1)^2+0.1(m-1)-0.055\bigr)
  (0.7+0.3e^{-2v}).
\end{aligned}
\end{equation}
Figure \ref{fig:factor-comparison} compares the proposed deformation functions with the empirical principal components identified by Cont and Vuleti\'c \cite{ContVuletic2023Simulation}.
The functions $f^1,\ldots,f^4$ are intended as tractable
approximations of the qualitative shapes of the empirical
components, rather than as estimates of the principal component
eigenfunctions themselves.

To incorporate these functions into a stochastic volatility surface,
we introduce amplitude parameters $\lambda_1,\ldots,\lambda_4\geq0$
and consider
\begin{gather}\label{eq:four-factor-model}
\sigma_t(m,v)
=
\bar\sigma
\exp\left(
\sum_{i=1}^4 \lambda_i f^i(m,v)Y_t^i
\right),
\end{gather}
where $\bar\sigma>0$ is a constant baseline volatility level and the noise
$
Y_t=(Y_t^1,\ldots,Y_t^4),
$
takes values in $[0,1]^4$.  The stochastic factors may, for example, be modeled by Jacobi
diffusions, however, the specific dynamics of $Y$
play no role in the static no-arbitrage analysis below. Thus, in the notation of Section~3, we have
$f^0=\log\bar\sigma$, while the four deterministic factor functions
are $\lambda_i f^i$, $i=1,\ldots,4$. The amplitude parameters control the magnitude of the deformations generated by the individual factors. We shall derive explicit restrictions on these parameters ensuring the absence of static arbitrage. Rather than checking the conditions of Lemma \ref{lem no static arbitrage}
directly, we apply the sufficient conditions established
in Propositions \ref{proposition Jacobi calendar}, \ref{prop suff cond for no call arb Jacobi} and \ref{Jacobi sufficient butterfly}. Our problem is to determine a nontrivial set of amplitude
vectors $\lambda\in[0,\infty)^4$ for which all three sufficient conditions 
hold. The following result provides such restrictions.

\begin{prop}
\label{thm:four-factor-amplitudes}
Let $\bar{\sigma}>0$ and let $f^1,\ldots,f^4$
be the functions defined in \eqref{eq:four-factors}.
For $\lambda\in[0,\infty)^4$, set
\[
\begin{aligned}
R&:=1.2\lambda_1+0.4\lambda_2
   +1.06\lambda_3+0.145\lambda_4,\\
L&:=0.5\lambda_1+\lambda_2
   +0.15\lambda_3+0.9\lambda_4,\\
L_+&:=\lambda_2+0.15\lambda_3
   +0.9\lambda_4,
\end{aligned}
\]
and denote
\[
\sigma_{\max}:=\bar{\sigma}e^R,\qquad
a:=\frac{\log(1.4)}
        {\bar{\sigma}\sqrt{30/365}}
  +\frac{\sigma_{\max}}{2},
\qquad
K=-\log(0.6).
\]
Suppose that
\begin{align}
0.84\lambda_1+0.28\lambda_2
  +0.087\lambda_4
&\leq\frac12,
\label{eq:amplitude-calendar}\\
L_+
&\leq\frac{1}{1.4\sigma_{\max}(1+a)},
\label{eq:amplitude-call}\\
2.8\left(K+\frac{\sigma_{\max}^2}{2}\right)L
+1.96\left(
\sigma_{\max}^2+K^2+\frac{\sigma_{\max}^4}{4}
\right)L^2
&\leq1.
\label{eq:amplitude-butterfly}
\end{align}
Conditions \eqref{eq:amplitude-calendar},
\eqref{eq:amplitude-call}, and
\eqref{eq:amplitude-butterfly} ensure the absence of
calendar spread, call spread, and butterfly arbitrage,
respectively. Then the four factor implied volatility model
\eqref{eq:four-factor-model} is free of static arbitrage
on $D$ at every time $t\geq0$.
\end{prop}

\medskip
Below we show that the admissible set of amplitudes contains a
four-dimensional cube with nonempty interior. Thus, all four deformation
patterns can be active simultaneously without introducing static arbitrage
on the prescribed domain.

\begin{cor}
Let $\bar{\sigma}=0.2$ and suppose that the amplitude parameters satisfy
\[
0\leq \lambda_i\leq 0.15,\qquad i=1,\ldots,4.
\]
Then the four factor implied volatility model \eqref{eq:four-factor-model}
is free of static arbitrage on $D$ at every time $t\geq0$.
\end{cor}

\noindent
{\bf Proof:} We verify the sufficient conditions of Proposition \ref{thm:four-factor-amplitudes}. Under the above
restrictions on the amplitudes,
\[
R\leq 0.42075,\qquad
L\leq 0.3825,\qquad
L_+\leq 0.3075.
\]
Consequently,
\[
\sigma_{\max}
=
\bar{\sigma}e^R
\leq
0.2e^{0.42075}
<0.305.
\]
The calendar spread condition \eqref{eq:amplitude-calendar} follows from
\[
\begin{aligned}
0.84\lambda_1+0.28\lambda_2+0.087\lambda_4
&\leq (0.84+0.28+0.087)0.15\\
&=0.18105<\frac12.
\end{aligned}
\]
For the call spread condition, we have
\[
a
=
\frac{\log(1.4)}
     {0.2\sqrt{30/365}}
+\frac{\sigma_{\max}}{2}
<6.1.
\]
Hence
\[
\frac{1}{1.4\sigma_{\max}(1+a)}
>
\frac{1}{1.4\cdot0.305\cdot7.1}
>0.33
>L_+,
\]
and \eqref{eq:amplitude-call} follows.

Finally, using $K=-\log(0.6)<0.511$, we obtain
\[
\begin{aligned}
&2.8\left(K+\frac{\sigma_{\max}^2}{2}\right)L
+1.96\left(
\sigma_{\max}^2+K^2+\frac{\sigma_{\max}^4}{4}
\right)L^2\\
&\quad<
2.8\left(0.511+\frac{0.305^2}{2}\right)0.3825\\
&\qquad
+1.96\left(
0.305^2+0.511^2+\frac{0.305^4}{4}
\right)0.3825^2\\
&\quad<0.70<1,
\end{aligned}
\]
which gives \eqref{eq:amplitude-butterfly}. Thus all three sufficient
conditions of Theorem~4.1 are satisfied, and the model is free of
static arbitrage on $D$ at every time $t\geq0$.
\hfill$\square$

The cube in Corollary~4.2 is only one simple example of a set of
admissible amplitude parameters. Other admissible regions can be
constructed directly from the three inequalities
\eqref{eq:amplitude-calendar}--\eqref{eq:amplitude-butterfly}.
In particular, decreasing the admissible range of one amplitude may
allow the ranges of other amplitudes to be enlarged.

\section{Proofs}\label{sec Proofs}

\subsection{Proof of Lemma \ref{lem no static arbitrage}}
Fix \(t\ge 0\). Since
$$
C_t(m,v)
=
S_t\cdot c^{BS}\bigl(m,v,\sigma_t(m,v)\bigr),
$$
and \(S_t>0\), the signs of the derivatives of \(C_t\) with respect to \(m\) and \(v\)
coincide with the signs of the corresponding derivatives of
$$
c(m,v):=c^{BS}\bigl(m,v,\sigma_t(m,v)\bigr).
$$
Therefore it is enough to compute the derivatives of \(c\). For simplicity, write
\[
\sigma=\sigma_t(m,v),
\qquad
d_1=d_1(\sigma,m,v),
\qquad
d_2=d_2(\sigma,m,v).
\]
A direct calculation gives
\[
c^{BS}_v
=
\frac{\sigma\varphi(d_1)}{2\sqrt v},
\qquad
c^{BS}_m
=
-\Phi(d_2),
\qquad
c^{BS}_\sigma
=
\sqrt v\,\varphi(d_1),
\]
and
\[
c^{BS}_{mm}
=
\frac{\varphi(d_2)}{m\sigma\sqrt v},
\qquad
c^{BS}_{m\sigma}
=
\frac{\varphi(d_2)d_1}{\sigma},
\qquad
c^{BS}_{\sigma\sigma}
=
\varphi(d_1)
\left(
\frac{d_1^2\sqrt v}{\sigma}
-
d_1v
\right).
\]
Consequently,
\[
\partial_v c
=
c^{BS}_v
+
c^{BS}_\sigma\,\partial_v\sigma, \qquad
\partial_m c
=
c^{BS}_m
+
c^{BS}_\sigma\,\partial_m\sigma,
\]
and
\[
\partial_{mm}c
=
c^{BS}_{mm}
+
2c^{BS}_{m\sigma}\partial_m\sigma
+
c^{BS}_{\sigma\sigma}(\partial_m\sigma)^2
+
c^{BS}_\sigma\partial_{mm}\sigma.
\]

Hence
\[
\partial_v c
=
\varphi(d_1)
\left(
\frac{\sigma}{2\sqrt v}
+
\sqrt v\,\partial_v\sigma
\right),
\]
and therefore
\begin{gather}\label{equiv 1}
\partial_v c\ge 0
\quad\Longleftrightarrow\quad
\frac{\sigma}{2v}
+
\partial_v\sigma
\ge0
\quad\Longleftrightarrow\quad
\partial_v \log\sigma+\frac{1}{2v}\geq 0.
\end{gather}

Similarly,
\begin{gather}\label{pochodna c po m w dowodzie}
\partial_m c
=
-\Phi(d_2)
+
\sqrt v\,\varphi(d_1)\partial_m\sigma.
\end{gather}
We shall use here the elementary identity
\begin{gather}\label{zwiazek miedzy varphi(d1) i varphi(d2)}
\varphi(d_2)=\frac1m\varphi(d_1).
\end{gather}
Indeed, since \(d_2=d_1-\sigma\sqrt v\), we have
\[
d_1^2-d_2^2
=
(d_1-d_2)(d_1+d_2).
\]
Moreover,
\[
d_1-d_2=\sigma\sqrt v
\]
and
\[
d_1+d_2=2d_1-\sigma\sqrt v
=
-\frac{2\log m}{\sigma\sqrt v}.
\]
Therefore
\[
d_1^2-d_2^2=-2\log m.
\]
Hence
\[
\frac{\varphi(d_2)}{\varphi(d_1)}
=
\exp\left(\frac{d_1^2-d_2^2}{2}\right)
=
\exp(-\log m)
=
\frac1m.
\]
Using \eqref{zwiazek miedzy varphi(d1) i varphi(d2)} in \eqref{pochodna c po m w dowodzie} we get
\begin{gather}\label{equiv 2}
\partial_m c\le0
\quad\Longleftrightarrow\quad
\partial_m\sigma
-
\frac{\Phi(d_2)}
{m\sqrt v\,\varphi(d_2)}
\le0.
\end{gather}
Dividing the right side by $\sigma$ and using the definition \eqref{H funkcja definicja} of the function $H_{m,v}$
we obtain 
\begin{gather}\label{equiv 2 prime}
\partial_m c\le0
\quad\Longleftrightarrow\quad
\partial_m\log\sigma
\leq H_{m,v}(\log \sigma).
\end{gather}

Finally,
\[
\partial_{mm}c
=
\frac{\varphi(d_2)}{m\sigma\sqrt v}
+
\frac{2d_1\varphi(d_2)}{\sigma}\partial_m\sigma
+
\varphi(d_1)
\left(
\frac{d_1^2\sqrt v}{\sigma}
-
d_1v
\right)
(\partial_m\sigma)^2
+
\sqrt v\,\varphi(d_1)\partial_{mm}\sigma.
\]
Again using \eqref{zwiazek miedzy varphi(d1) i varphi(d2)}
and dividing by \(\sqrt v\,\varphi(d_1)\), we obtain that
\begin{gather}\label{equiv 3}
\partial_{mm}c\ge0 \quad
\Longleftrightarrow \quad 
\partial_{mm}\sigma
+
\frac{d_1d_2}{\sigma}(\partial_m\sigma)^2
+
\frac{2d_1}{m\sigma\sqrt v}\partial_m\sigma
+
\frac{1}{m^2\sigma v}
\ge0.
\end{gather}
Substituting the identities 
\begin{gather}
\partial_m\sigma=\sigma \ \partial_m \log\sigma,
\qquad
\partial_{mm}\sigma
=\sigma\bigl(\partial_{mm}\log\sigma+(\partial_m \log\sigma)^2\bigr).
\end{gather}
into \eqref{equiv 3} and dividing by $\sigma>0$, we obtain
\begin{gather}\label{equiv 4}
\partial_{mm}c\ge0 \quad
\Longleftrightarrow \quad \partial_{mm}\log\sigma+(1+d_1d_2)(\partial_m \log\sigma)^2+
\frac{2d_1}{m\sigma\sqrt{v}}\partial_m \log\sigma+\frac{1}{m^2\sigma^2 v}\geq 0.
\end{gather}
By the definition of \(P_{\sigma,m,v}\), this is equivalent to
\[
\partial_{mm}\log\sigma
+
P_{\sigma,m,v}(\partial_m\log \sigma)
\ge0.
\]

Applying the pointwise equivalences \eqref{equiv 1}, \eqref{equiv 2 prime}, \eqref{equiv 4} for all \((m,v)\in D_t\), and then taking probabilities, gives that conditions \eqref{no calendar spread def}, \eqref{no call spread def}, 
\eqref{no butterfly spread def} are equivalent to \eqref{no calendar spread}, \eqref{no call spread}, 
\eqref{no butterfly spread}, respectively. \hfill$\square$

\subsection{Proof of Theorem \ref{tw o prawd. statycznego arbitrazu}}
By definition,
$
\sigma_t(m,v)=e^{F_t},
$
and, consequently, 
\[
\partial_v\log\sigma_t(m,v)=V_t,
\qquad
\partial_m\log\sigma_t(m,v)=M_t,
\qquad
\partial_{mm}\log\sigma_t(m,v)=K_t.
\]

We first consider calendar spread arbitrage. By Lemma \ref{lem no static arbitrage}, calendar spread arbitrage at the point \((m,v)\) occurs on the event
\[
\left\{
V_t<-\frac1{2v}
\right\}.
\]
Since \(V_t\) is Gaussian and \(\operatorname{Var}(V_t)>0\), we obtain
\[
\mathbb P
\left(
V_t<-\frac1{2v}
\right)
=
\Phi\left(
\frac{
-\frac1{2v}-\mathbb E[V_t]
}
{
\sqrt{\operatorname{Var}(V_t)}
}
\right).
\]
This proves (i).

We now consider call spread arbitrage. By Lemma \ref{lem no static arbitrage}, the call spread arbitrage at \((m,v)\) occurs on the event
$$
\left\{M_t
>H_{m,v}(F_t)\right\}.
$$
Since we omit the dependence of $F_t$ and $M_t$ on $(m,v)$, we also write $H$ instead of $H_{m,v}$ in the sequel. Thus the probability of occurrence of the call spread arbitrage can be computed by conditioning on $F_t$, that is 
\[
\mathbb P(M_t>H(F_t))
=
\int_{\mathbb R}
\mathbb P(M_t>H(y)\mid F_t=y)f_F(y)\,dy.
\]
Since \((F_t,M_t)\) is Gaussian, the conditional distribution of \(M_t\) given
\(F_t=y\) is normal with mean \(\mu_{M|F}(y)\) and variance
\(\sigma_{M|F}^2\). Hence
\[
\mathbb P(M_t>H(y)\mid F_t=y)
=
1-
\Phi\left(
\frac{
H(y)-\mu_{M|F}(y)
}
{
\sigma_{M|F}
}
\right).
\]
This proves (ii).

Finally, we consider butterfly arbitrage. 
By Lemma \ref{lem no static arbitrage}, butterfly arbitrage at \((m,v)\) occurs on the event
\[
\left\{K_t+P_{e^{F_t},m,v}(M_t)<0 \right\}.
\]
Conditioning on the vector \((F_t,M_t)\), we obtain
\[
\mathbb P
\left(
K_t<-P_{e^{F_t},m,v}(M_t)
\right)
\]
\[
=
\int_{\mathbb R}\int_{\mathbb R}
\mathbb P
\left(
K_t<-P_{e^y,m,v}(z)
\mid F_t=y,M_t=z
\right)
f_{M|F=y}(z)f_F(y)\,dz\,dy.
\]
Since \((F_t,M_t,K_t)\) is Gaussian, the conditional distribution of \(K_t\)
given \((F_t=y,M_t=z)\) is normal with mean
\(\mu_{K|F,M}(y,z)\) and variance \(\sigma_{K|F,M}^2\). Therefore
\[
\mathbb P
\left(
K_t<-P_{e^y,m,v}(z)
\mid F_t=y,M_t=z
\right)
=
\Phi\left(
\frac{
-P_{e^y,m,v}(z)-\mu_{K|F,M}(y,z)
}
{
\sigma_{K|F,M}
}
\right).
\]
Substituting this expression into the conditional integral gives the formula in (iii).
\hfill$\square$

\subsection{Proof of Proposition \ref{proposition Jacobi calendar}}
Fix \(t\ge0\), \(m>0\) and \(v>0\). By Lemma~1.2, the model is free of
calendar spread arbitrage at the point \((m,v)\) and time \(t\) if and only if
\begin{gather}\label{cal aux w dowodzie}
\mathbb P\left(
f_v^0(m,v)+\sum_{i=1}^n Y_t^i\cdot f_v^i(m,v)
\ge
-\frac1{2v}
\right)=1.
\end{gather}
Under the assumption \ref{full support}, the support of the random vector
$
Y_t=(Y_t^1,\ldots,Y_t^n)
$
coincides with the whole cube \([0,1]^n\) for any $t>0$. Therefore \eqref{cal aux w dowodzie} holds if and only if 
\[
f_v^0(m,v)+\sum_{i=1}^n y_i \cdot f_v^i(m,v)
\ge
-\frac1{2v}
\]
for every
$
y=(y_1,\ldots,y_n)\in[0,1]^n.
$
Clearly,
\[
\min_{y\in[0,1]^n}
\sum_{i=1}^n y_i \cdot f_v^i(m,v)
=
-\sum_{i=1}^n \bigl(f_v^i(m,v)\bigr)^-,
\]
where
$
a^-:=\max(-a,0).
$
\hfill$\square$

\subsection{Proof of Proposition \ref{prop no call spread Jacobi}}
Fix \(t\ge0\), \(m>0\) and \(v>0\). By Lemma \ref{lem no static arbitrage} , the model is free of call
spread arbitrage at the point \((m,v)\) and time \(t\) if and only if
\begin{gather*}
\mathbb{P}\left(
\partial_m\log\sigma_t(m,v)
\leq H_{m,v}\bigl(\log\sigma_t(m,v)\bigr)
\right)=1.
\end{gather*}
Define two functions on the cube $[0,1]^n$:
\[
F(y)
:=
f^0(m,v)+\sum_{i=1}^n y_i\cdot f^i(m,v),
\qquad
M(y)
:=
f^0_m(m,v)+\sum_{i=1}^n y_i \cdot f^i_m(m,v),
\]
with $y=(y_1,\ldots,y_n)\in[0,1]^n$. Then
\[
\sigma_t(m,v)=e^{F(Y_t)},
\qquad
\partial_m\log\sigma_t(m,v)=M(Y_t),
\]
and the call spread condition is equivalent to
\begin{gather}\label{no call arb dowod}
\mathbb P\left(
M(Y_t)\le H_{m,v}(F(Y_t))
\right)=1.
\end{gather}
By \eqref{full support}, condition \eqref{no call arb dowod} hold, for $t>0$, if and only if
\begin{gather}\label{no call arb 2 dowod}
M(y)\le H_{m,v}(F(y)),
\qquad
y\in[0,1]^n.
\end{gather}
We give a characterization of \eqref{no call arb 2 dowod} which is based on the geometrical approach for describing the positivity condition 
\begin{gather}\label{aux positivity call arb Jacobi}
H_{m,v}(F(y))-M(y)\geq 0, \qquad y\in[0,1]^n.
\end{gather}
Consider the affine mapping
\[
y \longmapsto (F(y),M(y)),
\qquad y\in[0,1]^n.
\]
Its image is a compact convex set in $\mathbb{R}^2$.
Condition \eqref{aux positivity call arb Jacobi} is equivalent to
\[
H_{m,v}(z)-w\geq 0
\]
for every point $(z,w)$ in this image.
Since the left-hand side is decreasing in $w$,
it suffices to verify this inequality on the
upper boundary of the image. We now show that this reduces the problem to
a collection of one-dimensional tests.

We begin with proving two simple auxiliary results which allow us to reduce the $n$-dimensional problem \eqref{aux positivity call arb Jacobi} to one dimension. To this end we consider the function 
\[
\Psi(y):=H_{m,v}(c+\langle a,y\rangle)-d-\langle b,y\rangle,
\qquad y\in[0,1]^n,
\]
with \(a,b\in\mathbb R^n\), \(c,d\in\mathbb R\).

\begin{lem}
The following assertions are equivalent:
\begin{enumerate}
\item
\[
\Psi(y)\ge0
\qquad\text{for all }y\in[0,1]^n.
\]
\item
\begin{gather}\label{I definicja}
H_{m,v}(c+z)-d-B(z)\ge0
\qquad\text{for all }z\in I:=
\Big[
-\sum_{i=1}^n a_i^-,
\sum_{i=1}^n a_i^+
\Big],
\end{gather}
where 
\begin{gather}\label{funkcja B definicja}
B(z)
:=
\max
\left\{
\langle b,y\rangle:
y\in[0,1]^n,\;
\langle a,y\rangle=z
\right\}.
\end{gather}
\end{enumerate}
\end{lem}
\noindent
{\bf Proof:}
Note that $I$ is the range of the function $y\mapsto \langle a,y\rangle$ 
on the cube $[0,1]^n$. Fix \(z\in I\). For every \(y\in[0,1]^n\) satisfying
\[
\langle a,y\rangle=z,
\]
we have
\[
\Psi(y)=H_{m,v}(c+z)-d-\langle b,y\rangle.
\]
Therefore the smallest possible value of \(\Psi\) on this level set is obtained by
maximizing \(\langle b,y\rangle\). Hence
\[
\inf_{\substack{y\in[0,1]^n\\ \langle a,y\rangle=z}}
\Psi(y)
=
H_{m,v}(c+z)-d-B(z).
\]
Taking the minimum over all admissible values of \(z\) proves the claim.
\hfill$\square$

\vskip2ex

We now describe the representation of the function \(B\) in \eqref{funkcja B definicja} based on a geometrical viewpoint. Let us consider the linear map from $[0,1]^n$ to $\mathbb{R}^2$ given by
$$
y\longmapsto (\langle a,y\rangle,\langle b,y\rangle).
$$
Its image is a convex polygon $\mathcal{P}\subset \mathbb{R}^2$ with vertices 
\begin{gather}\label{A_v i B_v def}
A_u:=\langle a,u\rangle,
\qquad
B_u:=\langle b,u\rangle,
\end{gather}
where $u:=(u_1,u_2,...,u_n)\in\{0,1\}^n$. Hence
\[
\mathcal P
:=
\operatorname{conv}
\left\{
(A_u,B_u):
u\in\{0,1\}^n
\right\}.
\]
Observe that
\[
B(z)=\max\{w:(z,w)\in \mathcal{P}\},
\]
so the graph of \(B\) coincides with the set
\[
\Gamma:=\{(z,w)\in \mathcal{P}:w=B(z)\}.
\]
Geometrically, \(\Gamma\) is the upper envelope of the polygon \(\mathcal{P}\).
This leads to the following representation of the function $B$.

\begin{lem}\label{lemma aux 2}
Assume that \(a\neq0\) and, for each vertex $u$ of the cube $[0,1]^n$, define $A_u, B_u$ by \eqref{A_v i B_v def}. Then there exist an integer
$
N\in\{2,\ldots,2^n\}
$
and vertices
$
u^{(1)},\ldots,u^{(N)}
\in\{0,1\}^n
$
such that, setting
\[
z_j:=A_{u^{(j)}},
\qquad
w_j:=B_{u^{(j)}},
\qquad
j=1,\ldots,N,
\]
one has
$
z_1<z_2<\cdots<z_N,
$
and
\[
\Gamma
=
\bigcup_{j=1}^{N-1}
[(z_j,w_j),(z_{j+1},w_{j+1})].
\]
Moreover, $I$ in \eqref{I definicja} has the form
$
I=[z_1,z_N],
$
and for every \(j=1,\ldots,N-1\),
\[
B(z)=\alpha_j+\beta_j z,
\qquad
z\in[z_j,z_{j+1}],
\]
where
\[
\beta_j
:=
\frac{w_{j+1}-w_j}
     {z_{j+1}-z_j},
\]
and
\[
\alpha_j
:=
w_j-\beta_j z_j.
\]
In particular, \(B\) is concave and piecewise affine on $I$.
\end{lem}

\noindent
{\bf Proof:} The set \(\Gamma\) consists of those points of \(\mathcal{P}\) whose second coordinate
is maximal among all points of \(\mathcal{P}\) with the same first coordinate.
Since \(\mathcal{P}\) is a convex polygon, the set \(\Gamma\) is composed of finitely many
line segments joining certain vertices of \(\mathcal{P}\). Let
\[
(z_1,w_1),\ldots,(z_N,w_N)
\]
be these vertices ordered by increasing first coordinate. Every vertex of \(\mathcal{P}\) is the image of at least one vertex of the cube. Consequently, 
\(\mathcal{P}\) has at most $2^n$ vertices, hence
$
N\le 2^n.
$
By construction,
$
I=[z_1,z_N].
$
On every segment joining
$
(z_j,w_j)
$
and
$
(z_{j+1},w_{j+1}),
$
the function \(B\) is affine. Its slope equals
\[
\beta_j
=
\frac{w_{j+1}-w_j}
     {z_{j+1}-z_j},
\]
and therefore
\[
B(z)
=
w_j+
\beta_j
(z-z_j).
\]
After rearranging terms,
\[
B(z)=\alpha_j+\beta_j z,
\]
where
\[
\alpha_j=w_j-\beta_j z_j.
\]
The concavity of \(B\) follows from the convexity of $\mathcal{P}$.
\hfill$\square$
\vskip2ex

Combining the previous two lemmas, we obtain the following characterization of the positivity of $\Psi(\cdot)$.
Note that if \(a=0\), then \(I=\{0\}\) and the problem reduces to checking the single inequality
\[
H_{m,v}(c)-d-\sum_{i=1}^n b_i^+\ge0.
\]
In what follows we assume \(a\neq0\).

\begin{cor}\label{cor pomocnicze call spread}
Let $z_1,...,z_N$, $\alpha_j,\beta_j$, $j=1,2,...,N-1$ be as in Lemma \ref{lemma aux 2}. Then the inequality
\begin{gather}\label{positivity of psi}
\Psi(y)\ge0,
\qquad y\in[0,1]^n,
\end{gather}
holds if and only if for every \(j=1,\ldots,N-1\),
\[
H_{m,v}(c+z)-d-\alpha_j-\beta_j z
\ge0
\qquad
\text{for all }z\in[z_j,z_{j+1}].
\]
Equivalently, it is sufficient and necessary for \eqref{positivity of psi} to verify:
\begin{enumerate}
\item \label{pierwszy warunek na polygon}
\[
H_{m,v}(c+z_j)-d-w_j\ge0,
\qquad j=1,\ldots,N;
\]
\item \label{drugi warunek na polygon} for every solution of
\[
H_{m,v}^{\prime}(c+z)=\beta_j, \quad z\in(z_j,z_{j+1}),\quad j=1,...,N-1
\]
one has
\[
H_{m,v}(c+z)-d-\alpha_j-\beta_j z\ge0.
\]
\end{enumerate}
\end{cor}

By Corollary \ref{cor pomocnicze call spread} one can reduce the problem of testing the positivity of $\Psi$ on the cube
$[0,1]^n$ to checking the positivity of $\Psi$ on the upper envelope of the polygon $\mathcal{P}\subset\mathbb{R}^2$. Indeed, \eqref{pierwszy warunek na polygon} corresponds to the vertices of the upper envelope and 
\eqref{drugi warunek na polygon} to line segments joining these vertices. 

To prove Proposition \ref{prop no call spread Jacobi} we apply this criterion with 
\begin{gather*}\label{notacja 1}
a_i:=f^i(m,v),
\qquad
b_i:=f_m^i(m,v),\qquad i=1,2,...,n,
\end{gather*}
together with
\begin{gather*}\label{notacja 2}
c:=f^0(m,v),
\qquad
d:=f_m^0(m,v).
\end{gather*}
Conditions characterizing exclusion of the call spread arbitrage follow directly from Corollary \ref{cor pomocnicze call spread}.

\subsection{Proof of Proposition \ref{prop suff cond for no call arb Jacobi}:} 
For every \(y=(y_1,\ldots,y_n)\in[0,1]^n\),
\[
\sum_{i=1}^n y_i\cdot f^i(m,v)
\le
\sum_{i=1}^n \bigl(f^i(m,v)\bigr)^+ .
\]
Since the function $H_{m,v}$ is strictly decreasing, see Lemma \ref{lem H decreasing} below, we obtain
\[
H_{m,v}\left(
f^0(m,v)+\sum_{i=1}^n y_i\cdot  f^i(m,v)
\right)
\ge
H_{m,v}\left(
f^0(m,v)+\sum_{i=1}^n \bigl(f^i(m,v)\bigr)^+
\right).
\]
Moreover,
\[
\sum_{i=1}^n y_i \cdot f^i_m(m,v)
\le
\sum_{i=1}^n \bigl(f^i_m(m,v)\bigr)^+ .
\]
Hence
\[
f^0_m(m,v)
+
\sum_{i=1}^n y_i \cdot f^i_m(m,v)
\le
f^0_m(m,v)
+
\sum_{i=1}^n \bigl(f^i_m(m,v)\bigr)^+ .
\]
Therefore
\[
M(y)\le H_{m,v}(F(y))
\]
for every \(y\in[0,1]^n\), which proves the claim.
\hfill$\square$

\begin{lem}\label{lem H decreasing}
For every fixed $m,v>0$, the function
$H_{m,v}:\mathbb{R}\to(0,\infty)$ is strictly decreasing.
\end{lem}
\noindent
{\bf Proof:} Recall that
\[
H_{m,v}(z)
=
\frac{\Phi(d_2(e^z,m,v))}
{me^z\sqrt{v}\,\varphi(d_2(e^z,m,v))},
\]
where
\[
d_2(\sigma,m,v)
=
-\frac{\log m}{\sigma\sqrt{v}}
-\frac{\sigma\sqrt{v}}{2}.
\]
Put $s=e^z\sqrt{v}$. Then
\[
d_2(e^z,m,v)
=
-\frac{\log m}{s}-\frac{s}{2}.
\]
Using the identity
\[
\frac{\Phi(x)}{\varphi(x)}=\int_{0}^{+\infty}\frac{\varphi(x-t)}{\varphi(x)}\,dt
=
\int_0^\infty
\exp\left(xt-\frac{t^2}{2}\right)\,dt,
\qquad x\in\mathbb{R},
\]
we obtain
\[
H_{m,v}(z)
=
\frac{1}{ms}
\int_0^\infty
\exp\left(
-\frac{\log m}{s}t
-\frac{s}{2}t
-\frac{t^2}{2}
\right)\,dt.
\]
Making the substitution $t=su$ gives
\[
H_{m,v}(z)
=
\frac{1}{m}
\int_0^\infty
\exp\left(
-u\log m
-\frac{s^2}{2}(u+u^2)
\right)\,du.
\]
Since $s^2=ve^{2z}$, we conclude that
\[
H_{m,v}(z)
=
\frac{1}{m}
\int_0^\infty
\exp\left(
-u\log m
-\frac{ve^{2z}}{2}(u+u^2)
\right)\,du.
\]
For every $u>0$, the integrand is strictly decreasing
as a function of $z$. Hence $H_{m,v}$ is strictly
decreasing on $\mathbb{R}$.
\hfill$\square$

\subsection{Proof of Proposition \ref{prop no butterfy arb Jacobi}}
Fix \(t\ge0\), \(m>0\) and \(v>0\). Define
\[
F(y)
:=
f^0(m,v)+\sum_{i=1}^n y_i\cdot  f^i(m,v),\qquad 
M(y)
:=
f^0_m(m,v)+\sum_{i=1}^n y_i\cdot  f^i_m(m,v),
\]
and
\[
K(y)
:=
f^0_{mm}(m,v)+\sum_{i=1}^n y_i\cdot f^i_{mm}(m,v),
\qquad
y\in[0,1]^n.
\]
Furthermore, let
\[
R(y)
:=
K(y)+P_{e^{F(y)},m,v}\bigl(M(y)\bigr),
\]
where $P_{\sigma,m,v}(\cdot)$ is the quadratic function given by \eqref{P funkcja kwadratowa}. By Lemma \ref{lem no static arbitrage} the model is free of butterfly arbitrage at the point
\((m,v)\) and time \(t\) if and only if
\[
\mathbb P\bigl(R(Y_t)\ge0\bigr)=1,
\]
which means that 
\[
R(y)\ge0,
\qquad
y\in[0,1]^n.
\]
The function $R$ depends on the factors \(y\) only through the
three affine functions \(F\), \(M\), and \(K\). This observation allows us to
reformulate the problem in geometric terms. We now reduce the butterfly condition to a condition on the projection of the polytope $Q$ defined in \eqref{def Q polytope}.  Note that, for \(y\in[0,1]^n\),
\[
F(y)=c+\langle a,y\rangle,
\qquad
M(y)=d+\langle b,y\rangle,
\qquad
K(y)=e+\langle r,y\rangle.
\]
Hence the butterfly condition is equivalent to
\[
e+\langle r,y\rangle
+
P_{e^{c+\langle a,y\rangle},m,v}
\bigl(d+\langle b,y\rangle\bigr)
\ge0,
\qquad
y\in[0,1]^n.
\]
Let
\[
D=\{(\langle a,y\rangle,\langle b,y\rangle):y\in[0,1]^n\}.
\]
For \((z,w)\in D\), define
\[
L(z,w)
=
\min\{\langle r,y\rangle:
y\in[0,1]^n,\ \langle a,y\rangle=z,\ \langle b,y\rangle=w\}.
\]
For fixed \((z,w)\in D\), the term
\[
P_{e^{c+z},m,v}(d+w)
\]
does not depend on the choice of \(y\) satisfying
$
\langle a,y\rangle=z,
\langle b,y\rangle=w.
$
Therefore the smallest possible value of the left-hand side of the butterfly
condition over this level set is
\[
e+L(z,w)+P_{e^{c+z},m,v}(d+w).
\]
Consequently, the butterfly condition holds for all \(y\in[0,1]^n\) if and
only if
\[
e+L(z,w)+P_{e^{c+z},m,v}(d+w)\ge0,
\qquad
(z,w)\in D.
\]

It remains to describe $L$. Since the cube $[0,1]^n$ is the
convex hull of its vertices, its image under the linear mapping
\[
y \longmapsto
\bigl(\langle a,y\rangle,\langle b,y\rangle,\langle r,y\rangle\bigr)
\]
is precisely
\[
Q=\operatorname{conv}
\bigl\{(A_q,B_q,C_q):q\in\{0,1\}^n\bigr\}.
\]
Consequently,
\[
L(z,w)=\min\{u\in\mathbb R:(z,w,u)\in Q\}.
\]
Thus, the graph of $L$ is the lower boundary of $Q$.
Since $Q$ is a convex polytope, $L$ is piecewise affine on $D$.

By the linear independence of $a$ and $b$, the set $D$ has
nonempty interior. Hence, we can choose a finite triangulation
\[
D=\bigcup_{k=1}^N T_k
\]
such that \(L\) is affine on each triangle \(T_k\). Thus, on
\(T_k\),
\[
L(z,w)=\alpha_k+\beta_k z+\gamma_k w.
\]
Therefore the condition
\[
e+L(z,w)+P_{e^{c+z},m,v}(d+w)\ge0,
\qquad
(z,w)\in D,
\]
is equivalent to
\[
e+\alpha_k+\beta_k z+\gamma_k w
+
P_{e^{c+z},m,v}(d+w)
\ge0
\]
for all \((z,w)\in T_k\) and all \(k=1,\ldots,N\). This proves the result.
\hfill$\square$
\subsection{ Proof of Proposition \ref{Jacobi sufficient butterfly}:}
For every \(y\in[0,1]^n\),
\[
\langle r,y\rangle
\ge
-\sum_{i=1}^n r_i^-.
\]
Moreover,
\[
c+\langle a,y\rangle
\in
\left[
c-\sum_{i=1}^n a_i^-,
\;
c+\sum_{i=1}^n a_i^+
\right],
\]
and
\[
d+\langle b,y\rangle
\in
\left[
d-\sum_{i=1}^n b_i^-,
\;
d+\sum_{i=1}^n b_i^+
\right].
\]
Therefore
\[
P_{e^{c+\langle a,y\rangle},m,v}
\bigl(d+\langle b,y\rangle\bigr)
\ge
\inf_{\substack{
u\in
[c-\sum_i a_i^-,
 \;c+\sum_i a_i^+]\\
w\in
[d-\sum_i b_i^-,
 \;d+\sum_i b_i^+]
}}
P_{e^u,m,v}(w).
\]
 Combining the above inequalities yields
\[
e+\langle r,y\rangle
+
P_{e^{c+\langle a,y\rangle},m,v}
\bigl(d+\langle b,y\rangle\bigr)
\ge0,
\]
which proves the claim.\hfill$\square$

\subsection{Proof of Proposition \ref{PP}}

Write
\[
g_t(m,v):=\log\sigma_t(m,v)
=a(v)-\lambda\rho(m)+r(v)Y_t.
\]
Since $Y_t\in[0,1]$ almost surely, we have
\[
\partial_v g_t(m,v)
=a'(v)+r'(v)Y_t\geq 0,
\]
and
\[
\partial_m g_t(m,v)
=-\lambda\rho'(m)\leq 0.
\]
Consequently,
\[
\partial_v g_t(m,v)+\frac{1}{2v}\geq 0
\]
and
\[
\partial_m g_t(m,v)
\leq H_{m,v}(g_t(m,v)),
\]
since $H_{m,v}$ is positive. Thus the no calendar
spread and no call spread conditions in Lemma \ref{lem no static arbitrage} hold.

It remains to verify the no butterfly arbitrage
condition. We have
\[
\partial_{mm}g_t(m,v)
=-\lambda\rho''(m)\geq 0.
\]
Therefore it suffices to show that
\[
P_{\sigma_t(m,v),m,v}
\bigl(-\lambda\rho'(m)\bigr)\geq 0
\]
for all $(m,v)\in D_t$, almost surely. Consider the function
\[
G(\lambda,m,v,y)
:=
P_{e^{a(v)-\lambda\rho(m)+r(v)y},m,v}
\bigl(-\lambda\rho'(m)\bigr).
\]
For $\lambda=0$,
\[
G(0,m,v,y)
=\frac{1}{m^2e^{2a(v)+2r(v)y}v}>0.
\]
Since $D_t\times[0,1]$ is compact,
\[
\min_{(m,v,y)\in D_t\times[0,1]}
G(0,m,v,y)>0.
\]
By continuity of $G$, there exists $\lambda_0>0$
such that
\[
G(\lambda,m,v,y)\geq0
\]
for every $\lambda\in[0,\lambda_0]$,
$(m,v)\in D_t$ and $y\in[0,1]$. In particular,
\[
P_{\sigma_t(m,v),m,v}
\bigl(\partial_m g_t(m,v)\bigr)\geq0
\]
almost surely, uniformly on $D_t$. Combining this
with $\partial_{mm}g_t(m,v)\geq0$, we obtain
\[
\partial_{mm}g_t(m,v)
+
P_{\sigma_t(m,v),m,v}
\bigl(\partial_m g_t(m,v)\bigr)\geq0.
\]
By Lemma \ref{lem no static arbitrage}, the no butterfly arbitrage condition
holds. This completes the proof.
\hfill$\square$

\subsection{Proof of Proposition \ref{thm:four-factor-amplitudes}}
\noindent
The model \eqref{eq:four-factor-model} is a particular
case of \eqref{model with bounded factors 1}  with
\[
f^0(m,v)=\log\bar{\sigma}
\]
and factor functions $\lambda_i f^i$, $i=1,\ldots,4$.
We verify successively the sufficient conditions of
Propositions \ref{proposition Jacobi calendar}, \ref{prop suff cond for no call arb Jacobi} and \ref{Jacobi sufficient butterfly}.

\medskip
\noindent
\textit{Calendar spread.}
Differentiating the functions in
\eqref{eq:four-factors} with respect to $v$, we obtain
\[
\begin{aligned}
f^1_v
 &=-0.7e^{-2v}(1-0.5(m-1)),\\
f^2_v
 &=-0.7(m-1)e^{-2v},\\
f^3_v
 &=5\operatorname{sech}^2(5(v-0.45))
       (1+0.15(m-1)),\\
f^4_v
 &=-0.6e^{-2v}
       ((m-1)^2+0.1(m-1)-0.055).
\end{aligned}
\]
Since $m\in[0.6,1.4]$ and $v>0$,
\[
(f^1_v)^-\leq0.84,\qquad
(f^2_v)^-\leq0.28,\qquad
(f^3_v)^-=0,\qquad
(f^4_v)^-\leq0.087.
\]
Consequently, by \eqref{eq:amplitude-calendar}, 
\[
\sum_{i=1}^4(\lambda_i f^i_v)^-
\leq
0.84\lambda_1+0.28\lambda_2
+0.087\lambda_4
\leq\frac12.
\]
Since $f^0_v=0$ and $v\leq1$, we have
$1/(2v)\geq1/2$. Hence
\[
\sum_{i=1}^4(\lambda_i f^i_v)^-
\leq \frac{1}{2v}+f^0_v,
\]
and Proposition \ref{proposition Jacobi calendar} yields the absence of calendar
spread arbitrage.

\medskip
\noindent
\textit{Call spread.}
Let
\[
A(v):=0.65+0.35e^{-2v},\qquad
B(v):=0.7+0.3e^{-2v},
\]
and
\[
T(v):=\tanh(5(v-0.45)).
\]
The derivatives with respect to $m$ are
\[
\begin{aligned}
f^1_m&=-0.5A(v),&
f^2_m&=A(v),\\
f^3_m&=0.15T(v),&
f^4_m&=(2(m-1)+0.1)B(v).
\end{aligned}
\]
Since $A(v),B(v)\leq1$ and $|T(v)|\leq1$,
\[
\sum_{i=1}^4(\lambda_i f^i_m)^+
\leq L_+.
\]
The bounds
\[
|f^1|\leq1.2,\qquad
|f^2|\leq0.4,\qquad
|f^3|\leq1.06,\qquad
|f^4|\leq0.145
\]
give
\[
\log\bar{\sigma}
\leq
\log\bar{\sigma}
+\sum_{i=1}^4(\lambda_i f^i)^+
\leq
\log\bar{\sigma}+R.
\]
Set
\[
z=\log\bar{\sigma}
+\sum_{i=1}^4(\lambda_i f^i(m,v))^+,
\qquad s=e^z.
\]
Then
\[
\bar{\sigma}\leq s\leq\sigma_{\max}.
\]
Recall that
\[
d_2(s,m,v)
=-\frac{\log m}{s\sqrt v}
 -\frac{s\sqrt v}{2}.
\]
Since $m\leq1.4$, $v\geq30/365$, $v\leq1$, and
$\bar{\sigma}\leq s\leq\sigma_{\max}$,
\[
d_2(s,m,v)
\geq
-\frac{\log(1.4)}
       {\bar{\sigma}\sqrt{30/365}}
-\frac{\sigma_{\max}}{2}
=-a.
\]
The function $x\mapsto\Phi(x)/\varphi(x)$ is increasing,
and the standard Mills ratio inequality gives
\[
\frac{\Phi(-a)}{\varphi(-a)}
\geq\frac{1}{1+a}.
\]
It follows that
\[
\begin{aligned}
H_{m,v}\left(
\log\bar{\sigma}
+\sum_{i=1}^4(\lambda_i f^i)^+
\right)
&\geq
\frac{1}{1.4\sigma_{\max}(1+a)}\\
&\geq L_+\\
&\geq
\sum_{i=1}^4(\lambda_i f^i_m)^+.
\end{aligned}
\]
Since $f^0_m=0$, Proposition \ref{prop suff cond for no call arb Jacobi} yields the absence
of call spread arbitrage.

\medskip
\noindent
\textit{Butterfly spread.}
The second derivatives with respect to $m$ satisfy
\[
f^1_{mm}=f^2_{mm}=f^3_{mm}=0,
\qquad
f^4_{mm}=2B(v)>0.
\]
Hence
\[
\sum_{i=1}^4(\lambda_i f^i_{mm})^-=0.
\]
Moreover,
\[
\sum_{i=1}^4 |\lambda_i f^i_m| \leq L,
\qquad
\sum_{i=1}^4 |\lambda_i f^i| \leq R.
\]
For fixed $(m,v)\in D$, consider the intervals appearing
in Proposition \ref{Jacobi sufficient butterfly}:
\[
I_u:=
\left[
\log\bar{\sigma}
-\sum_{i=1}^4(\lambda_i f^i)^-,
\,
\log\bar{\sigma}
+\sum_{i=1}^4(\lambda_i f^i)^+
\right]
\]
and
\[
I_w:=
\left[
-\sum_{i=1}^4(\lambda_i f^i_m)^-,
\,
\sum_{i=1}^4(\lambda_i f^i_m)^+
\right].
\]
The above estimates imply that
\[
I_w\subset[-L,L],
\]
while, for every $u\in I_u$,
\[
e^u
\leq
\bar{\sigma}
\exp\left(
\sum_{i=1}^4 |\lambda_i f^i|
\right)
\leq
\bar{\sigma}e^R
=
\sigma_{\max}.
\]
Set $\sigma=e^u$. Recall that
\[
P_{\sigma,m,v}(w)
=
(1+d_1d_2)w^2
+\frac{2d_1}{m\sigma\sqrt v}\,w
+\frac{1}{m^2\sigma^2v}.
\]
A direct calculation yields
\[
\begin{aligned}
m^2\sigma^2vP_{\sigma,m,v}(w)
={}&1+
2m\left(-\log m+\frac{\sigma^2v}{2}\right)w\\
&+
m^2\left(
\sigma^2v+(\log m)^2
-\frac{\sigma^4v^2}{4}
\right)w^2.
\end{aligned}
\]
For $u\in I_u$ and $w\in I_w$, we have
$\sigma=e^u\leq\sigma_{\max}$ and $|w|\leq L$.
Since $m\leq1.4$, $|\log m|\leq K$, and $v\leq1$,
we therefore obtain
\[
\begin{aligned}
m^2\sigma^2vP_{\sigma,m,v}(w)
\geq{}&
1-2.8\left(K+\frac{\sigma_{\max}^2}{2}\right)L\\
&-1.96\left(
\sigma_{\max}^2+K^2+\frac{\sigma_{\max}^4}{4}
\right)L^2.
\end{aligned}
\]
By \eqref{eq:amplitude-butterfly}, the right-hand side
is nonnegative. Hence
\[
P_{e^u,m,v}(w)\geq0,
\qquad
u\in I_u,\quad w\in I_w.
\]
Since
\[
f^0_{mm}=0,
\qquad
\sum_{i=1}^4(\lambda_i f^i_{mm})^-=0,
\]
the sufficient condition of Proposition \ref{Jacobi sufficient butterfly} is satisfied,
and therefore butterfly arbitrage is excluded.
\hfill$\square$

\bibliographystyle{amsplain}
\bibliography{implvol}

\end{document}